\documentclass[prd,showpacs,preprintnumbers,groupedaddress,superscriptaddress,nofootinbib,11pt]{revtex4-1} 

\usepackage{amsthm,amsmath}
\newtheoremstyle{newplain}
{}
{}
{\it}
{}
{\bfseries}
{.}
{5pt}
{\thmname{#1}\hspace{5pt}\thmnumber{#2}\thmnote{\hspace{2pt}[{\normalfont #3}]}}
\theoremstyle{newplain}

\newtheorem{proposition}{Proposition}
\usepackage{cases}
\usepackage{ulem}
\usepackage{amsfonts}
\usepackage{mathrsfs}

\usepackage[dvipdfmx]{graphicx}
\usepackage[hang,small,bf]{caption}
\usepackage[subrefformat=parens]{subcaption}
\usepackage{afterpage}
\usepackage{here}

\usepackage{hyperref} 
\hypersetup{%
 setpagesize=false,
 bookmarksnumbered=true,%
 bookmarksopen=true,%
 colorlinks=true,%
 linkcolor=blue,%
 citecolor=red}

\usepackage{color}
\usepackage{comment}

\newcommand{\tcal}[1]{\widetilde{\mathcal{#1}}}

\begin{document}

\preprint{YITP-25-22}
\preprint{RUP-25-2}

\title{Shadow formation in gravitational collapse: Redshift and blueshift by spacetime dynamics
}



\author{Yasutaka Koga}
\affiliation{Department of Information and Computer Science, Osaka Institute of Technology, Hirakata 573-0196, Japan}
\affiliation{Center for Gravitational Physics and Quantum Information, Yukawa Institute for Theoretical Physics, Kyoto University, Kyoto 606-8502, Japan}

\author{Nobuyuki Asaka}
\affiliation{TDSE Inc., Tokyo 163-1427, Japan}

\author{Masashi Kimura}
\affiliation{Department of Information, Artificial Intelligence and Data Science, Daiichi Institute of Technology, Tokyo 110-0005, Japan}
\affiliation{Department of Physics, Rikkyo University, Tokyo 171-8501, Japan}

\author{Kazumasa Okabayashi}
\affiliation{Center for Gravitational Physics and Quantum Information, Yukawa Institute for Theoretical Physics, Kyoto University, Kyoto 606-8502, Japan}



\date{\today}

\begin{abstract}
A black hole illuminated by a background light source is observed as a black hole shadow.
For a black hole formed by gravitational collapse of a transmissive object, redshift of light due to the spacetime dynamics is expected to play a crucial role in the shadow formation.
In this paper, we investigate the redshift of light caused by the spacetime dynamics.
First, we consider a spherical shell model.
We see that the collapse of the shell typically leads to the redshift of light, while blueshift can be also observed in some cases.
This result suggests that a shadow image is generally formed in the late stage of the gravitational collapse of a transmissive object.
Second, we propose a covariant formula for the redshift of light in a general, dynamical, and spherically symmetric spacetime.
This formula relates the redshift to the energy-momentum tensor of the background spacetime and provides its intuitive interpretation with a Newtonian analogy.
The redshift effect analyzed in this work is regarded as the integrated Sachs-Wolfe effect or the Rees-Sciama effect in gravitational collapse.
\end{abstract}


\maketitle


\tableofcontents

\section{Introduction}
\label{sec:introduction}
A black hole accompanied with a light source around or behind it is observed as a dark image in optical observation.
The dark image is called a black hole shadow and is considered to provide a key evidence of a black hole and the geometrical information in the vicinity.
In 2019, the Event Horizon Telescope (EHT) Collaboration established the first shadow image of the super massive black hole of M87 and succeeded in constraining the mass and spin~\cite{EventHorizonTelescope:2019dse}.
Subsequently, they also showed the shadow of the black hole at the center of Sgr. A* in 2022~\cite{EventHorizonTelescope:2022wkp}.
Although precise determination of the astrophysical parameters of black holes is still challenging, we will be able to access more information about them in the near future.

Depending on the light source configuration, the formation of a black hole shadow involves several mechanisms; absence of light emission from the horizon, accumulation of photons on the photon sphere, redshift of light by the gravitational potential, and so on.
The simplest theoretical model that reflects the spacetime structure would be a black hole illuminated by a distant 
spherical bright screen surrounding both the observer and the black hole~\cite{Gralla_2019}.
In this case, the emergence of the shadow for a stationary black hole can be understood as a consequence of no light emission from the white hole.
The apparent size of the shadow corresponds to that of the photon sphere, i.e., the radius of circular photon orbits, because it gives the threshold for light rays from infinity and the horizon~\cite{AKPande_1986}.
Since this description does not depend on the spacetime symmetries, this is also true for a dynamical but eternal black hole, such as the Vaidya spacetime~\cite{Koga:2022dsu}. 
Moreover, this description for the black hole shadow allows us to show that the observer can never see the merger of black hole shadows in the head-on coalescence of two black holes~\cite{Okabayashi_2020}.
See Refs.~\cite{Synge:1966okc,Claudel:2000yi,Shiromizu:2017ego,Yoshino:2017gqv,Yoshino:2019dty,Koga:2019uqd,Koga:2020akc,Kobialko:2020vqf,Siino:2019vxh,Siino:2021kep,
Amo:2023ofn,Cunha:2018acu,Mishra_2019,Solanki_2022} and the citations therein for the extensive works on black hole shadow and photon spheres.

The above argument for the 
shadow image does not apply to a black hole dynamically formed by the gravitational collapse 
because there is no white hole in this case.
For example, a distant observer may observe light emission from any direction if the collapsing object is transmissive.
This does not mean that a shadow is not formed but a dark image should be formed due to redshift of light that passed through the collapsing object.
This redshift effect is caused by the spacetime dynamics and is also known as the integrated Sachs-Wolfe or the Rees-Sciama effect in the cosmological context~\cite{Sachs:1967er,Rees:1968zza}.

In this paper, we investigate the redshift of a light ray in dynamical spherically symmetric spacetime.
We assume that there is a gravitating object at the center which is transmissive and collapsing.
This background dynamics affects the energy of a light ray propagating inside the object.
We see how the dynamics of gravitational collapse is related to the redshift of light and try to understand the mechanism with physical intuition.

This paper is organized as follows.
In Sec.~\ref{sec:preliminary}, we clarify our motivation and focus.
In Sec.~\ref{sec:collapsing-shell}, we consider a spherical shell model and investigate the redshift of light in several cases.
We also explicitly show the shadow images formed during gravitational collapse of the shell in Sec.~\ref{sec:shadow-image}.
In Sec.~\ref{sec:generalspacetime}, we consider a general, dynamical, spherically symmetric spacetime.
We derive a new covariant formula for the redshift and discuss the physical interpretation.
In Sec.~\ref{sec:summary}, we summarize our work and make a conclusion.
We also discuss the applications and significance of our results.
We adopt the geometrical units, $c=G=1$, throughout this paper.
We use Roman indices $a,b,...$ for the abstract index notation of tensors~\cite{textbook:wald}.
For the contraction of metric tensors or energy-momentum tensors with vectors, we also use brackets.
For example, for a metric $g$ and vectors $u,v$, we denote $g(u,v)=g_{ab}u^av^b$.

\section{Motivation and Focus}
\label{sec:preliminary}
Our 
focus is the redshift of light due the spacetime dynamics.
In the following, we clarify our motivation 
and the definition of {\it the redshift factor} in the present context.

\subsection{BH shadow of an eternal black hole}
\label{sec:shadow-eternal}
Let us review the shadow formation in a static black hole spacetime.
As a simple case, we consider the Schwarzschild black hole illuminated by a distant bright screen which is spherical and infinitely wide so that it surrounds both the observer and the black hole~\cite{Gralla_2019}.
The observer is also distant from the black hole.
This setup is useful for understanding the role of asymptotic behavior of null geodesics in shadow formation.
\footnote{
One can consider different models for the light source setup, such as light emission from accreting matter onto a black hole.
This is the case of the black hole shadow image of the M87~\cite{EventHorizonTelescope:2019dse,EventHorizonTelescope:2022wkp}.
See also Ref.~\cite{Gralla_2019} and the references therein for other simple theoretical models.
}
For a distant observer to observe light rays from the bright source, there must be the corresponding null geodesics connecting the bright screen and the observer.
The behavior of null geodesics is analyzed by means of the effective potential.
The Schwarzschild spacetime is given by the metric,
\begin{align}
    g=-f(r)dt^2+f^{-1}(r)dr^2+r^2d\Omega^2,
    \;\;f(r)=1-2M/r.\nonumber
\end{align}
The null condition for the null geodesic tangent $k^\mu=dx^\mu/d\lambda$ is given by $g(k,k)=0$.
Using the conserved energy $E=-g(k,\partial_t)$ and conserved angular momentum $L=g(k,\partial_\phi)$ associated with the Killing vectors $\partial_t$ and $\partial_\phi$, respectively, and assuming that the orbit is confined in the equatorial plane $\theta=\pi/2$ without loss of generality, the equation reduces to the one-dimensional problem,
\begin{align}
    \dot r^2+V(b;r)=0,\;\; V(b;r):=-1+b^2f(r)r^{-2},
\end{align}
where $b:=L/E$ is the conserved impact parameter and we have normalized the affine parameter as $\lambda\to\lambda/E$.

\begin{figure}[h]
\centering
\includegraphics[width=200pt]{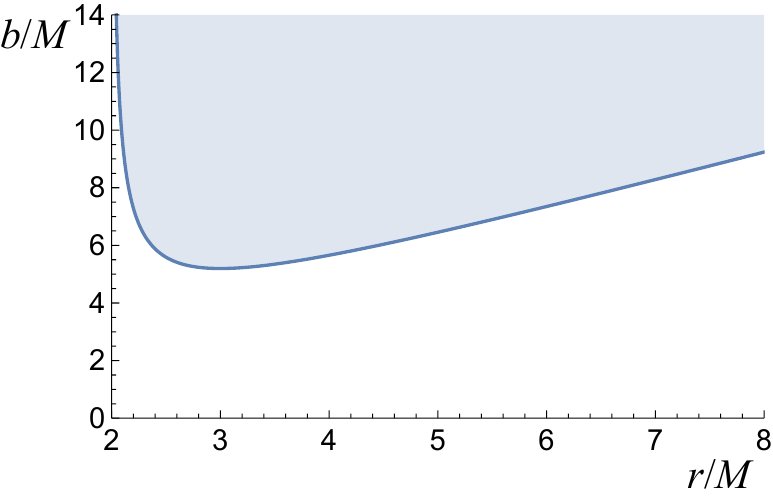}
\caption{\label{fig:Vplot}
The $r$-$b$ plane and the forbidden region $V(b;r)>0$ (the shaded region) for null geodesics in the Schwarzschild spacetime.
}
\end{figure}

In the $r-b$ plane in Fig.~\ref{fig:Vplot}, null geodesics are described as horizontal lines.
The null geodesics connecting the bright screen and the observer should be the orbits emitted from and escaping to infinity.
They correspond to the horizontal lines touching and reflected by the boundary of the forbidden region, $V(b;r)>0$.
These null geodesics have a supercritical impact parameter, $b>b_c$, where the critical value is given by $b_c=3\sqrt{3}M$ and the critical geodesics with $b=b_c$ are those that asymptote the photon sphere radius at $r=3M$.
Conversely, there are no incoming light rays with subcritical impact parameter, $b<b_c$.
Since the impact parameter is related to the incident angle, the absence of incoming subcritical light rays implies that the observer observes a dark image within the corresponding critical incident angle.

\begin{figure}[h]
\centering
\includegraphics[width=200pt]{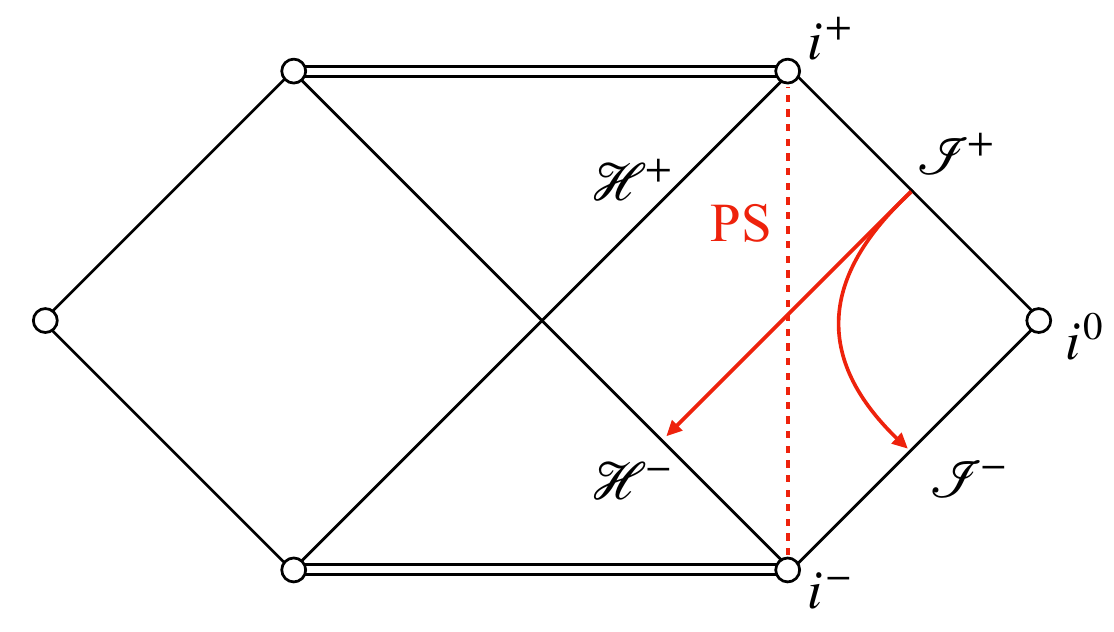}
\caption{\label{fig:raytracing-eternal}
Ray-tracing from $\mathscr{I}^+$ in an eternal black hole.
}
\end{figure}

Let us revisit the shadow imaging from the causal viewpoint.
As the distant observer and the distant light source are connected by null geodesics, we can regard that the observer is on the future null infinity $\mathscr{I}^+$ and the source is on the past null infinity $\mathscr{I}^-$.
What we actually do in the imaging is ray-tracing of null geodesics backward in time.
In the Schwarzschild spacetime (Fig.~\ref{fig:raytracing-eternal}), past-directed null geodesics from $\mathscr{I}^+$ are classified into three types.
Those going to $\mathscr{I}^-$, $\mathscr{H}^-$, and $i^-$.
They correspond to the supercritical $b>b_c$, subcritical $b<b_c$, and critical $b=b_c$ orbits, respectively, in the above potential analysis.
Therefore, the absence of incoming light rays to the observer within the critical incident angle is the consequence of the existence of $\mathscr{H}^-$, i.e., the white hole horizon.
In other words, one can say that the black hole shadow is the image of the white hole.
This statement would be true for any other spacetimes as far as it is an eternal black hole.
In fact, this is confirmed to be true in an eternal black hole case of the Vaidya spacetime, a black hole spacetime with accreting null dust~\cite{Koga:2022dsu}.

\subsection{BH shadow in gravitational collapse}
\begin{figure}[h]
\centering
\includegraphics[width=120pt]{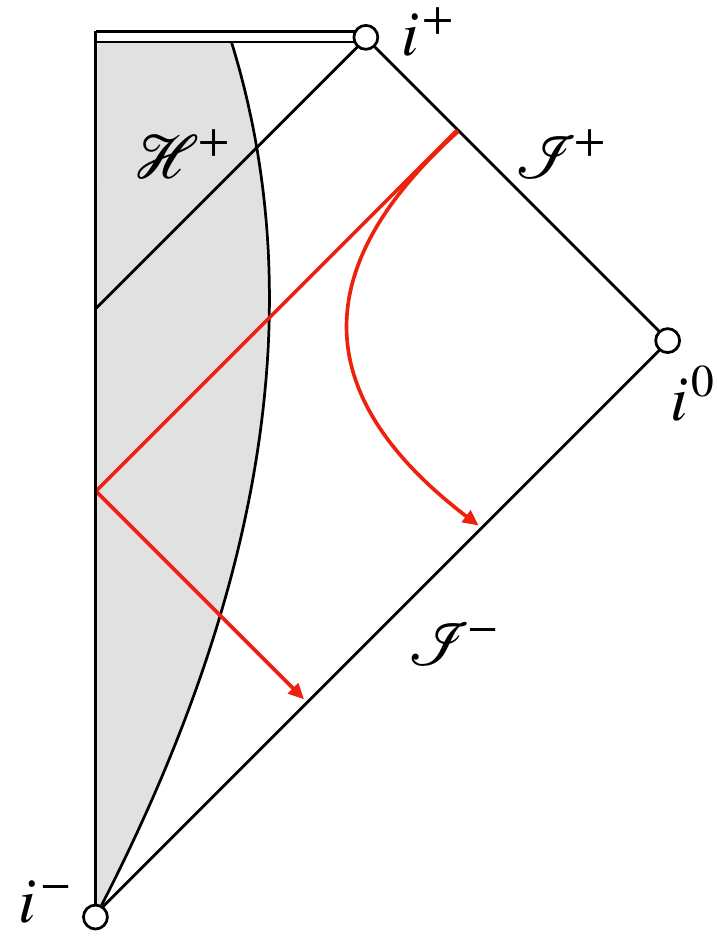}
\caption{\label{fig:raytracing-collapse}
Ray-tracing from $\mathscr{I}^+$ in a collapsing black hole.
}
\end{figure}

Let us consider a black hole dynamically formed by the gravitational collapse of a transmissive object.
The causal structure of the spacetime is described as Fig.~\ref{fig:raytracing-collapse}.
The key difference from the eternal black hole case is the absence of the white hole.
In this spacetime, any past-directed null geodesics from $\mathscr{I}^+$ go to $\mathscr{I}^-$, implying that the observer receives a light ray from any direction.
However, this does not necessarily mean that the observer never observes a black hole shadow.
Rather, it implies that the shadow should be formed due to the redshift of the light rays.

In this case, as we have assumed that there is no interaction between light and the collapsing object, light is redshifted only by the effect of the spacetime dynamics.
However, the relation between the spacetime dynamics and the redshift of light is quite nontrivial.
Moreover, light might be blueshifted even in gravitational collapse.
Our aim in the current paper is to understand how light is redshifted or blueshifted depending on the spacetime dynamics, especially in the case of gravitational collapse.

\subsection{Redshift and spacetime dynamics}
In this paper, we focus on an asymptotically flat spherically symmetric spacetime $(\mathcal{M},g)$ and null geodesics, or light rays, propagating from $\mathscr{I}^-$ to $\mathscr{I}^+$.
In the asymptotic region, the metric takes the form, \footnote{
Due to the spherical symmetry, we can choose the radial coordinate $r$ as the area radius. The time coordinate $t$ is chosen so that $\partial_t$ is perpendicular to $\partial_r$. Then, there are no degrees of freedom of the Lorentz transformations in this coordinate system in the asymptotic region if the spacetime is curved.}
\begin{align}
    g=\eta+\mathcal{O}(r^{-1}),\;\;\eta=-dt^2+dr^2+r^2d\Omega_2^2,\nonumber
\end{align}
and the time coordinate basis $\partial_t$ defines the asymptotic energy of a light ray as $E:=-g(k,\partial_t)$, where $k$ is the null geodesic tangent.
We define the redshift factor of a light ray as
\begin{align}
\label{eq:def-redshift}
\alpha:=\frac{E|_{\mathscr{I}^+}}{E|_{\mathscr{I}^-}},
\end{align}
and investigate its behavior under several dynamics of the spacetime.

We should clarify the difference between our definition of the redshift and that frequently appearing in the literature.
Usually, the terminology of ``(gravitational) redshift" is used for the redshift of light due to the static gravitational potential.
For example, suppose there is a point light source in the depth of the Schwarzschild spacetime and an observer receives the light ray at infinity.
Their unit four velocities are given as $u_\mathrm{src}=\sqrt{f(r_\mathrm{src})}^{-1}\partial_t$ and $u_\mathrm{obs}=\sqrt{f(r_\mathrm{obs})}^{-1}\partial_t\sim\partial_t$, respectively, where $f(r)=1-2M/r$ and $r_\mathrm{src}$ and $r_\mathrm{obs}\gg r_\mathrm{src}$ are their radii.
The redshift by the static gravitational field is given as the ratio of their proper energies,
\begin{align}
\alpha_\mathrm{static}=\frac{E_\mathrm{obs}^\mathrm{p}}{E_\mathrm{src}^\mathrm{p}}=\frac{-g(k,u_\mathrm{obs})}{-g(k,u_\mathrm{src})}.
\end{align}
Since the geometrical energy associated with the Killing vector $E=-g(k,\partial_t)$ is conserved, we have
\begin{align}
\alpha_\mathrm{static}=\sqrt{\frac{f(r_\mathrm{obs})}{f(r_\mathrm{src})}}<1.
\end{align}
The light emitted from the depth of the static potential is thus redshifted.

On the other hand, the redshift of our definition~\eqref{eq:def-redshift} is trivial for any null geodesics in the Schwarzschild spacetime.
That is, due to the conservation of the Killing energy in a static spacetime, we always have $\alpha=1$.
The redshift factor can deviate from unity only if the spacetime is dynamical and thus $\partial_t$ cannot be taken as the global timelike Killing vector.
Note that the redshift factor~\eqref{eq:def-redshift} can be also regarded as the ratio of the proper energies, where the source is put at infinity and $u_\mathrm{src}\sim \partial_t$.

In the following, we investigate the redshift factor~\eqref{eq:def-redshift} in several cases of a dynamical shell model and discuss its qualitative behavior in a generic spacetime by proposing a new covariant formula.

\section{Collapsing shell and redshift of light}
\label{sec:collapsing-shell}
In this section, we study a spherical shell model and see that light rays are typically redshifted if the shell is collapsing.
The behavior ensures that an observer at infinity observes shadow formation.

\subsection{Construction of spacetime}
We consider that spacetime $(\mathcal{M},g)$ is divided into two parts $(M_\pm,g_\pm)$ by a spherical shell, i.e., a timelike hypersurface $\Sigma$ with a spherical spacelike section.
Thanks to the Birkhoff's theorem, the outside region has the Schwarzschild geometry,
\begin{equation}
g_+=-f(r_+)dt^2+f^{-1}(r_+)dr_+^2+r_+^2d\Omega^2,
\end{equation}
where $f(r)=1-2M/r$, and the inside region has the Minkowski geometry,
\begin{equation}
g_-=-dT^2+dr_-^2+r_-^2d\Omega^2.
\end{equation}
These two regions are joined at a timelike hypersurface $\Sigma$ specified by
\begin{align}
    r_+=r_-=R(\tau),
\end{align}
where $R(\tau)$ is the shell motion and $\tau$ is the proper time of the shell.

We impose Israel's first junction condition.
The four-velocity of the shell is given by
\begin{equation}
u=\frac{dt}{d\tau}\partial_t+\frac{dr_+}{d\tau}\partial_r^+,
\end{equation}
in the Schwarzschild side.
We require the normalization,
\begin{equation}
\label{eq:normalization-sch}
-1=g_+(u,u)=-f(r)\left(\frac{dt}{d\tau}\right)^2+f^{-1}(r)\left(\frac{dr}{d\tau}\right)^2.
\end{equation}
The induced metric on $\Sigma$ in the Schwarzschild side is then
\begin{eqnarray}
\label{eq:h+}
h_+&=&-f(r_+)\left(\frac{dt}{d\tau}\right)^2d\tau^2+f^{-1}(r_+)\left(\frac{dr_+}{d\tau}\right)^2d\tau^2+r_+^2d\Omega^2\nonumber\\
&=&-d\tau^2+r_+^2d\Omega^2.
\end{eqnarray}
In the Minkowski side, we have
\begin{equation}
u=\frac{dT}{d\tau}\partial_T+\frac{dr_-}{d\tau}\partial_r^-
\end{equation}
and require the normalization,
\begin{equation}
\label{eq:normalization-min}
-1=g_-(u,u)=-\left(\frac{dT}{d\tau}\right)^2+\left(\frac{dr_-}{d\tau}\right)^2.
\end{equation}
The induced metric on $\Sigma_-$ is 
\begin{eqnarray}
\label{eq:h-}
h_-&=&-\left(\frac{dT}{d\tau}\right)^2d\tau^2+\left(\frac{dr_-}{d\tau}\right)^2d\tau^2+r_-^2d\Omega^2\nonumber\\
&=&-d\tau^2+r_-^2d\Omega^2.
\end{eqnarray}
Then, from Eqs.~\eqref{eq:h+} and~\eqref{eq:h-} and the fact $r_+=r_-$ on $\Sigma$, the first junction condition $h_+=h_-=: h$ is satisfied.
Hereafter we omit the sign $\pm$ of the areal radius $r_\pm$ as far as it does not matter.
\footnote{
Note that, since the areal radius $r:=r_\pm$ (at $p\in M_\pm$) is a $C^0$ function across $\Sigma$, it cannot be a smooth radial coordinate on $(\mathcal{M},g)$.
For example, this fact appears as the discontinuity of the coordinate bases $\partial_r^\pm$ in the following analysis.
}

The metric distribution is not smooth across the shell and the energy-momentum tensor $T$ is divergent at the hypersurface.
As the spacetime $(\mathcal{M},g)$ is vacuum except for the shell, we have
\begin{align}
    T_{ab}=\delta(l)S_{ab},
\end{align}
where $S$ is the surface stress-energy tensor of the shell.
The radial coordinate $l$ is the affine parameter of the spacelike geodesics perpendicularly penetrating the shell.
The value is taken as $l=0$ on the shell, $l>0$ in the exterior, and $l<0$ in the interior region, and is normalized so that $\nabla_a l=n_a$ for the unit normal vector $n$ to the shell.
$\delta(l)$ is the Dirac delta function.
According to Ref.~\cite{textbook:poisson}, $S$ is given as
\begin{align}
    S_{ab}=-\frac{1}{8\pi}\left([[\mathcal{K}_{ab}]]-[[\mathrm{tr}\mathcal{K}]]h_{ab}\right)
\end{align}
in terms of the extrinsic curvature $\mathcal{K}_{ab}=h_a{}^ch_b{}^d\nabla_cn_d$ and its trace $\mathrm{tr}\mathcal{K}=h^{ab}\mathcal{K}_{ab}$, where $[[X]]:=X_+-X_-$ denotes the jump of a quantity $X$ at the shell.
This equation is called the second junction condition.

As indicated in Appendix~\ref{sec:shell-em-tensor}, the extrinsic curvature is related to the shell dynamics $r=R(\tau)$.
In an ordinary procedure, by specifying the equation of state of the shell and restricting the surface stress tensor $S$, the dynamics of the shell $r=R(\tau)$ is determined through the second junction condition.
However, we specify the shell dynamics $R(\tau)$ by hand instead of specifying the equation of state.
The integration of Eqs.~\eqref{eq:normalization-sch} and~\eqref{eq:normalization-min} then gives $t(\tau)$ and $T(\tau)$, respectively, and the transformation between the time coordinates $t$ and $T$ at $\Sigma$ is obtained as $t(T)=t[\tau(T)]$, where $\tau(T)$ is the inverse function of $T(\tau)$.

\subsection{Light orbits and conserved energies}
We consider light orbits backward in time from an observer at $\mathscr{I}^+$ to the distant light source at $\mathscr{I}^-$.
The orbits are obtained by solving the null geodesic equation.
Here we assume the orbits are confined in the equatorial plane $\theta=\pi/2$ without loss of generality.

Let $\gamma(\lambda)$ be a null geodesic with its affine parameter $\lambda$ and $k$ be its tangent vector.
In the Schwarzschild region, the geodesic equation reduces to
\begin{align}
\label{eq:VS}
&\dot{r}^2+V(E,L;r)=0,
\;\;\; V(E,L;r):=-(E^2-L^2f(r)r^{-2}),
\end{align}
where $\dot{}=d/d\lambda$ and $E:=-g(\partial_t,k)=\text{const.}$ is the energy conserved in this region.
The angular momentum $L:=g(\partial_\phi,k)=\text{const.}$ is globally conserved.
If the null geodesic crosses the shell $\Sigma$ and enters the Minkowski region, the geodesic equation reduces to
\begin{align}
\label{eq:VM}
&\dot{r}^2+U(\mathcal{E},L;r)=0,\;\; U(\mathcal{E},L;r):=-(\mathcal{E}^2-L^2r^{-2}),
\end{align}
where $\mathcal{E}:=-g(\partial_T,k)=\text{const.}$ is the energy conserved in this region.
It can be integrated as
\begin{align}
\label{eq:Min-orbit}
-(T-T_*)^2+r^2=(L/\mathcal{E})^2.
\end{align}
The geodesic is extended across $\Sigma$ so that its tangent vector is continuous.
In fact, the geodesic equation,
\begin{equation}
\frac{dk^\mu}{d\lambda}+\Gamma^\mu_{\nu\rho}k^\nu k^\rho=0,
\end{equation}
in a coordinate system covering a neighborhood of $\Sigma$ implies that ${dk^\mu}/{d\lambda}$ is finite because the continuity of the metric guarantees the finiteness of $\Gamma^\mu_{\nu\rho}$.

Since the geodesic tangent $k$ is continuous across $\Sigma$, the energies $E=-g(\partial_t,k)$ and $\mathcal{E}=-g(\partial_T,k)$ are related to each other through the transformation of the coordinate basis.
The transformation $\{\partial_t,\partial_r^+\} \to \{\partial_T,\partial_r^-\}$ is given by\footnote{
This relation follows from the conditions in Eqs.~\eqref{eq:shell-4velocity} and \eqref{eq:shell-normal} that the four-velocities of the shell calculated on both sides $M_\pm$ are identical, as are the outward perpendicular vectors to the shell.
}
\begin{align}
\label{eq:coordtrans-t}
\partial_T
=&f^{-1}(r)\left[\sqrt{1+ R'^{2}}\sqrt{f(r)+R'^2}-R'^2\right]\partial_t
+\left[\sqrt{1+R'^2}-\sqrt{f(r)+R'^2}\right]R'\partial_r^+,\\
\label{eq:coordtrans-r}
\partial_r^-
=&f^{-1}(r)\left[\sqrt{1+R'^2} -\sqrt{f(r)+R'^2} \right]R'\partial_t
+\left[\sqrt{1+R'^2}\sqrt{f(r)+R'^2} -R'^2 \right]\partial_r^+,
\end{align}
where $R':=dR/d\tau$ and $\partial_r^+$ and $\partial_r^-$ are the radial coordinate basis for the Schwarzschild and Minkowski regions, respectively.
The basis $\partial_\theta$ and $\partial_\phi$ are not involved in the transformation as they are global coordinate basis of the spacetime.
Using the equation for the time basis, we obtain the transformation between the energies $E$ and $\mathcal{E}$ as
\begin{align}
\label{eq:etoE-general}
\mathcal{E}=f^{-1}(r)\left[\sqrt{1+R'^2}\sqrt{f(r)+R'^2}-R'^2\right]E
-f^{-1}(r)\left[\sqrt{1+R'^2}-\sqrt{f(r)+R'^2}\right]R' \dot{r}_+.
\end{align}

\subsection{Redshift in generic dynamics}
\begin{figure}[h]
\centering
\includegraphics[width=150pt]{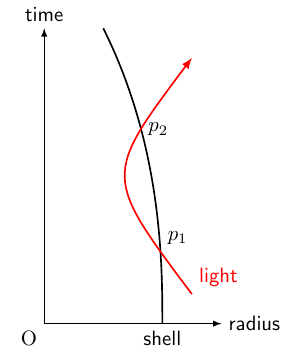}
\caption{\label{fig:shellandlight}
The schematic picture of the shell motion and the light orbit.
The light enters the shell at $p_1$ and exits at $p_2$.
}
\end{figure}

We analyze the redshift of light passing through the transmissive shell.
We suppose that the light ray enters the shell at the event $p_1$ and exits at the event $p_2$, as described in Fig.~\ref{fig:shellandlight}.
Let $E_1$ and $E_2$ be the conserved energies in the Schwarzschild side at the points $p_1$ and $p_2$, respectively.
For simplicity, we focus on the ratio $\alpha_{12}:=E_2/E_1$.
If the light ray goes to  the infinities, $\mathscr{I}^\pm$, without crossing the shell at other points, the ratio $\alpha_{12}$ coincides with the redshift factor~\eqref{eq:def-redshift} measured in the infinity.
\footnote{
In exotic dynamics of a shell, a light ray can cross the shell more than twice.
For example, in the case of ultracompact object formation, a light ray can cross the shell four times as shown in Fig.~\ref{fig:nulluco-orbits}.
Letting the events be $p_1,..., p_4$ in order, the redshift factor is obtained as the combination of the ratios, $\alpha=E_4/E_1=(E_2/E_1)(E_4/E_3)=\alpha_{12}\alpha_{34}$, where the conservation of energy in the Schwazschild region implies $E_3=E_2$.
}

Letting $W:=R'$ and using Eqs.~\eqref{eq:etoE-general} and~\eqref{eq:VS}, in general we have
\begin{align}
\label{eq:E/e}
\frac{\mathcal{E}}{E}=f^{-1}\left[
\left(\sqrt{1+W^2}\sqrt{f+W^2}-W^2\right)
-\sigma W\left(\sqrt{1+W^2}-\sqrt{f+W^2}\right)\sqrt{1-b^2f r^{-2}}
\right],
\end{align}
where $\sigma:=\mathrm{sign}(\dot{r})$, $b:=L/E$ is the conserved impact parameter in the  Schwarzschild region, and $f(r)$ is shortly denoted as $f$.
For the light ray crossing the shell twice, the redshift factor is obtained as
\begin{align}
\label{eq:alpha-e2/e1}
\alpha_{12}
&=\frac{E_2}{E_1}
=\frac{\mathcal{E}/E_1}{\mathcal{E}/E_2}\nonumber\\
&=\frac{f_2}{f_1}
\frac{\left(\sqrt{1+W_1^2}\sqrt{f_1+W_1^2}-W_1^2\right)
-\sigma_1 W_1\left(\sqrt{1+W_1^2}-\sqrt{f_1+W_1^2}\right)\sqrt{1-b_1^2f_1 r_1^{-2}}
}{\left(\sqrt{1+W_2^2}\sqrt{f_2+W_2^2}-W_2^2\right)
-\sigma_2 W_2\left(\sqrt{1+W_2^2}-\sqrt{f_2+W_2^2}\right)\sqrt{1-b_2^2f_2 r_2^{-2}}
},
\end{align}
where the subscripts ``1" and ``2" stand for values at the points $p_1$ and $p_2$, respectively.

$\mathcal{E}$ is the energy in the Minkowski side between the events $p_1$ and $p_2$.
Substituting the relation $b_1=L/E_1=(L/E_2)(E_2/E_1)=\alpha_{12} b_2$ into Eq. (\ref{eq:alpha-e2/e1}) and solving the equation about $\alpha_{12}$, we finally obtain
\begin{align}
\label{eq:redshift-formula}
\alpha_{12}=\frac{f_2}{f_1}\frac{
A_1\left(A_2-\sigma_2 C_2\sqrt{1-b_2^2f_2 r_2^{-2}}\right)\pm \sigma_1C_1\sqrt{\left(A_2-\sigma_2 C_2\sqrt{1-b_2^2f_2 r_2^{-2}}\right)^2-b_2^2f_2^2r_1^{-2}}
}{
\left(A_2-\sigma_2 C_2\sqrt{1-b_2^2f_2 r_2^{-2}}\right)^2+C_1^2b_2^2f_2^2 f_1^{-1}r_1^{-2}
},
\end{align}
where we have introduced the new variables,
\begin{align}
A_i&=\sqrt{1+W_i^2}\sqrt{f_i+W_i^2}-W_i^2,\\
C_i&=W_i\left(\sqrt{1+W_i^2}-\sqrt{f_i+W_i^2}\right),
\end{align}
which depend only on the radius $r_i$ and the shell's velocity $W_i$ at each event $p_i$ for $i=1,2$.
To summarize the formula~\eqref{eq:redshift-formula}, the redshift $\alpha_{12}$ is determined by the seven parameters $(r_1,r_2,W_1,W_2,\sigma_1,\sigma_2,b_2)$.
$r_i$ is the radius, $W_i=R'|_{p_i}$ is the shell's radial velocity, and $\sigma_i=\mathrm{sign}(\dot r)|_{p_i}=\pm 1$ indicates the radial direction of the light ray, inward($-$) or outward($+$), at each event $p_i$ for $i=1,2$ (see Fig.~\ref{fig:shellandlight} for description of our situation).
$b_2$ is the conserved impact parameter of the light ray at and after $p_2$, where the other impact parameter $b_1$ at and before $p_1$ is a redundant parameter obtained from the relation $b_2=\alpha_{12}b_1$.
The metric coefficient $f_i=f(r_i)$ is evaluated at each radius.
Note that we need to choose the sign $\pm$ in the numerator appropriately.
In the following, we focus on cases where $W_1=0$, and thus $C_1=0$, and do not need to specify the sign.
For the determination of the sign in generic cases, see Appendix~\ref{sec:redshift-pm}.

Note that the range of the parameters is restricted by physical consistency.
First, the orbital parameters $r_1,r_2,\sigma_1,\sigma_2,b_2$ should be consistent with the orbit in the interior Minkowski region, Eq.~\eqref{eq:Min-orbit}.
Second, the four velocities of the shell and the light should be specified consistently so that the light can enter and exit the shell at $p_1$ and $p_2$, respectively.
Third, the potential $V$ in Eq.~\eqref{eq:VS} should be negative.
These consistency conditions are summarized in Appendix~\ref{sec:consistency}.

\begin{figure}[h]
\centering
\includegraphics[width=150pt]{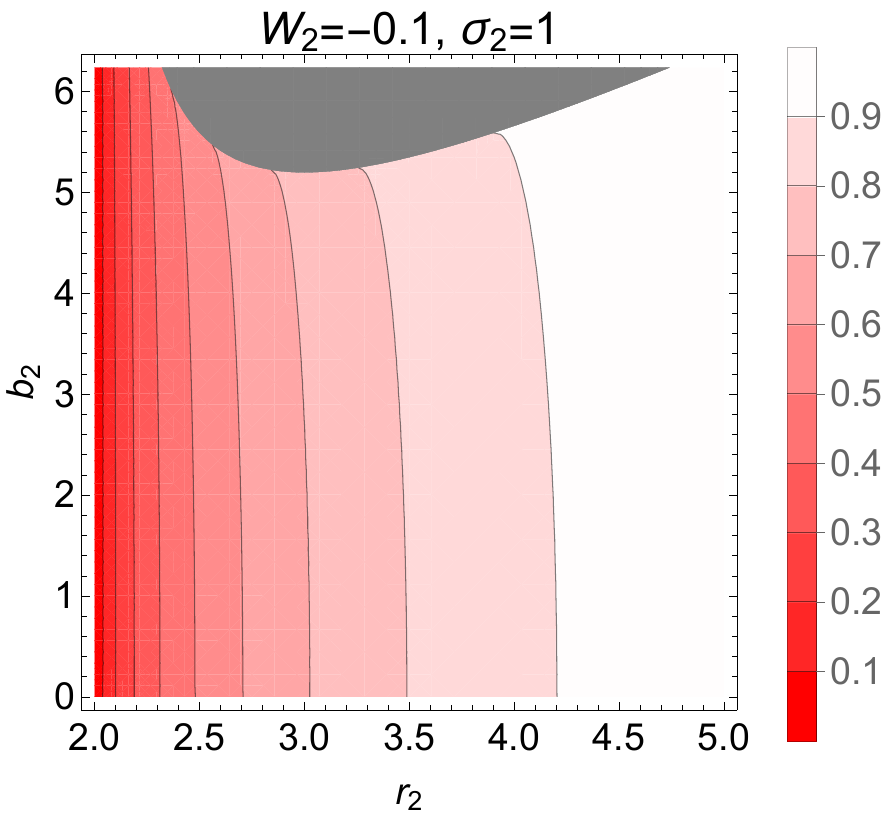}
\includegraphics[width=150pt]{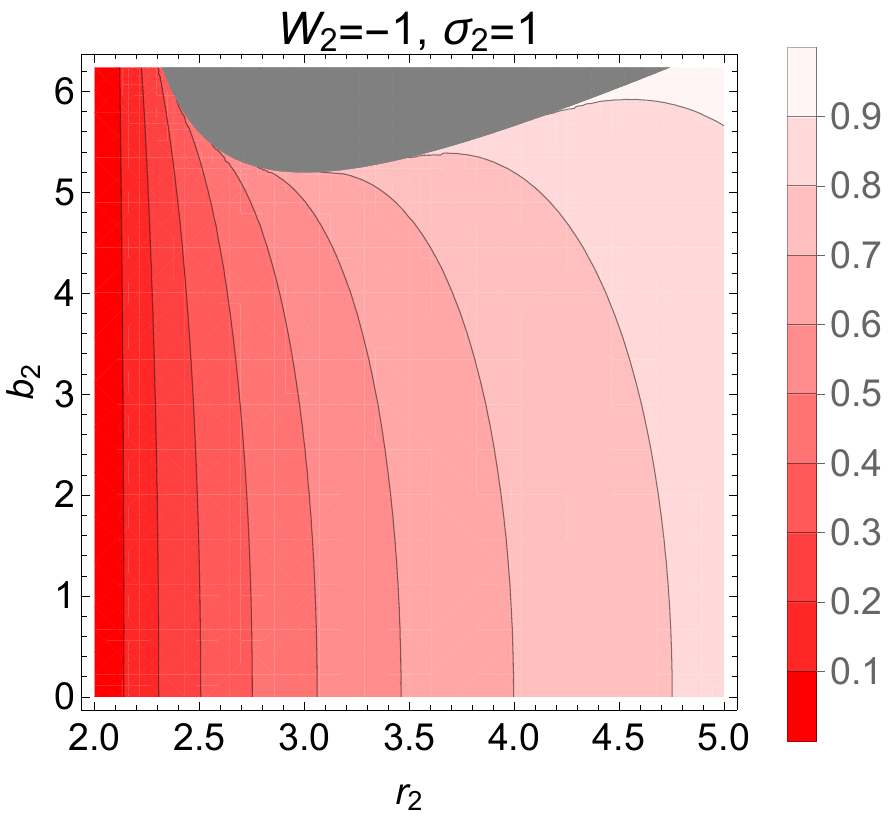}
\includegraphics[width=150pt]{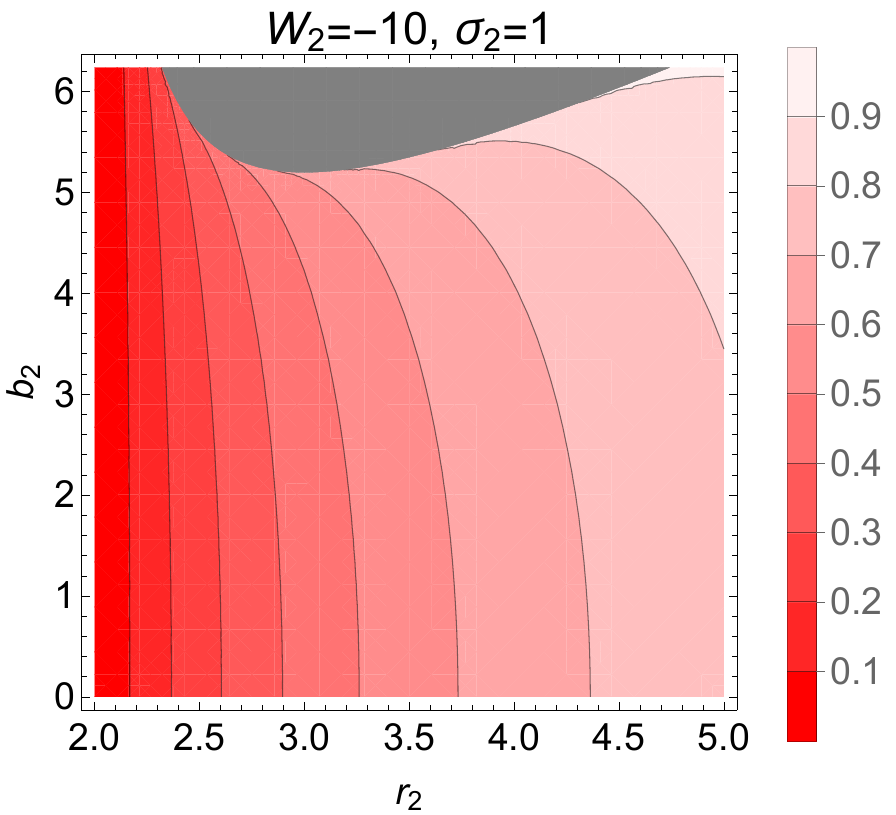}
\caption{\label{fig:collapse-outgoing}
Plot of the redshift factor $\alpha_{12}$ from Eq.~\eqref{eq:redshift-formula} of outgoing ($\sigma_2=+1$) lights from a collapsing shell with its velocity, $W_2=-0.1,\; -1,\; -10$.
The other parameters are fixed as $r_1=5M$, $W_1=0$, and $\sigma_1=-1$ with the unit fixed by the ADM mass $M=1$.
The shell is collapsing with the velocity $W_2=-0.1,\; -1,\; -10$ when the light exits.
The shaded region is forbidden by the consistency conditions~\eqref{eq:entry-condition}--\eqref{eq:nullvector-condition}.
}
\end{figure}

\begin{figure}[H]
\centering
\includegraphics[width=150pt]{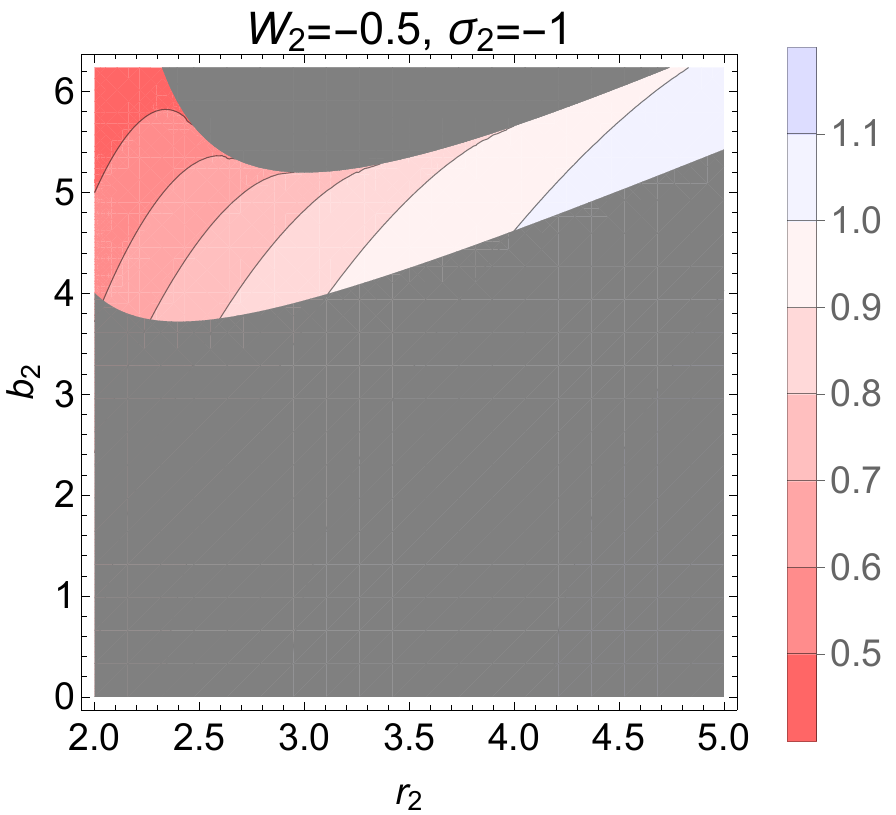}
\includegraphics[width=150pt]{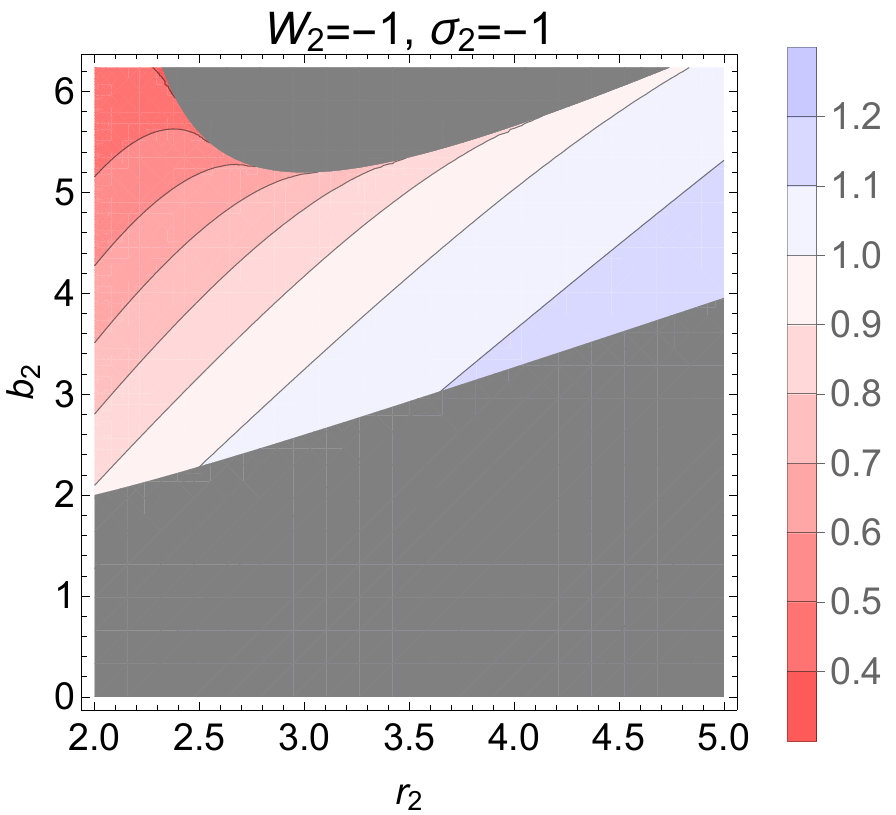}
\includegraphics[width=150pt]{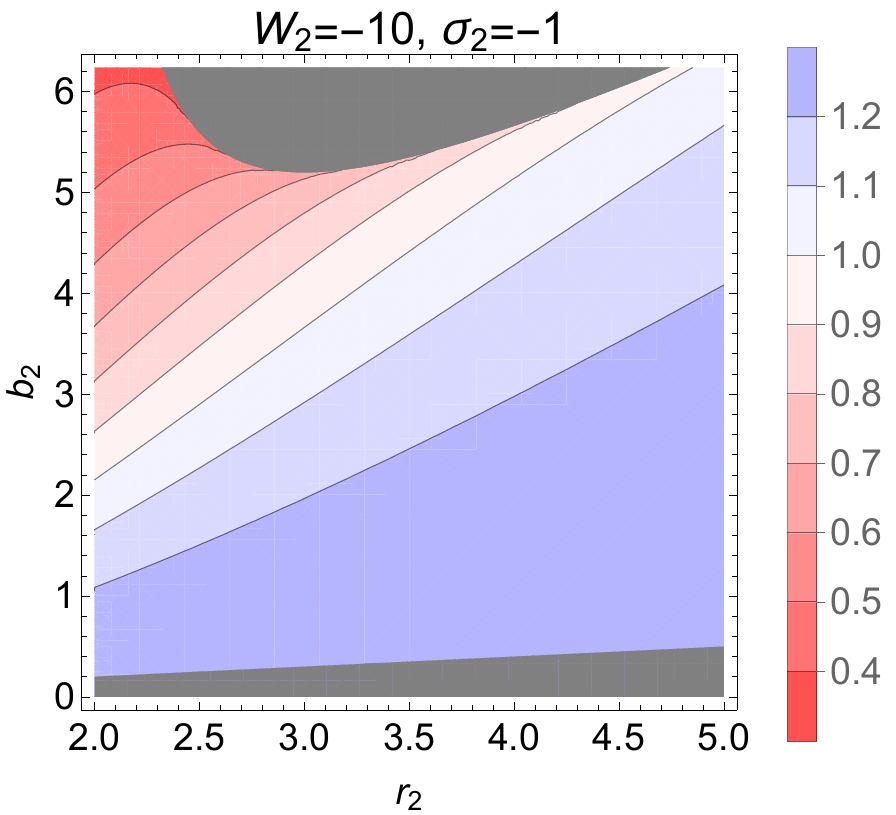}
\caption{\label{fig:collapse-ingoing}
Plot of the redshift factor $\alpha_{12}$ from Eq.~\eqref{eq:redshift-formula} of ingoing ($\sigma_2=-1$) lights from a collapsing shell with its velocity, $W_2=-0.5,\; -1,\; -10$.
The other parameters are $r_1=5M$, $W_1=0$, and $\sigma_1=-1$ with the unit $M=1$.
The shaded region is forbidden by the consistency conditions~\eqref{eq:entry-condition}--\eqref{eq:nullvector-condition}.
}
\end{figure}

Let us examine the redshift factor, Eq.~\eqref{eq:redshift-formula}, in some characteristic cases.
Suppose that the shell is static when a light ray enters the shell, $r_1=5M$ and $W_1=0$, and is shrinking when the light exits, $r_2<r_1$ and $W_2<0$.
This case requires $\sigma_1=-1$.
For some values of $W_2$, the redshift factor is obtained as Figs.~\ref{fig:collapse-outgoing} ($\sigma_2=+1$) and~\ref{fig:collapse-ingoing} ($\sigma_2=-1$).
We can see that the outgoing light rays ($\sigma_2=+1$) are definitely redshifted, $\alpha_{12}<1$.
If the shell is collapsing rapidly, i.e., $|W_2|$ is larger,  the redshift becomes more effective.
The strongest redshift $\alpha_{12}\to 0$ is achieved for arbitrary impact parameter near the horizon formation at $r=2M$, while such a light ray can escape to infinity if the impact parameter is subcritical, $b_2<b_c=3\sqrt{3}M$ (see Sec.~\ref{sec:shadow-eternal}).
This result indicates that the black hole shadow should be eventually formed in the end of gravitational collapse and its apparent size coincides with that of the photon sphere.
Moreover, rapid collapse leads to a darker image at the earlier stage of the collapse.

On the other hand, ingoing light rays ($\sigma_2=-1$) can be blueshifted if they exit at larger radius and have smaller impact parameters and if the shell is collapsing rapidly.
If the shell is going to collapse into a black hole, many of the blueshifted light rays will be swallowed.
However, some of them can escape to infinity if they are reflected by the effective potential of the Schwarzschild region, whose maximum is at the photon sphere $r=3M$.
This condition corresponds to $r_2>3M$ and $b_2>b_c$.

\begin{figure}[h]
\centering
\includegraphics[width=150pt]{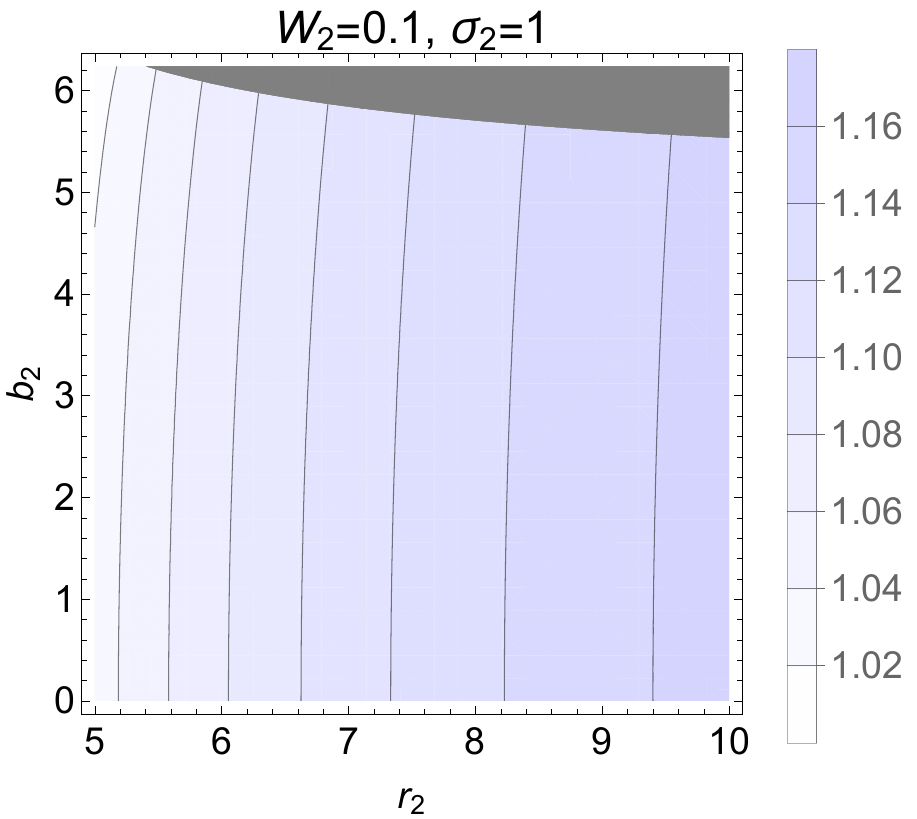}
\includegraphics[width=150pt]{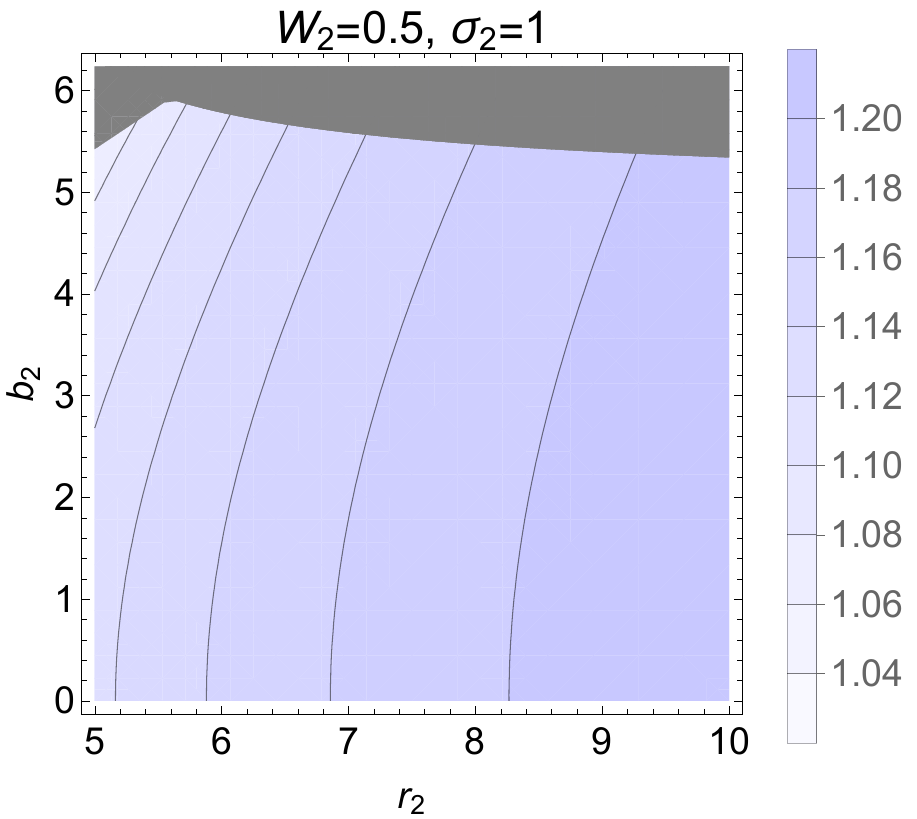}
\includegraphics[width=150pt]{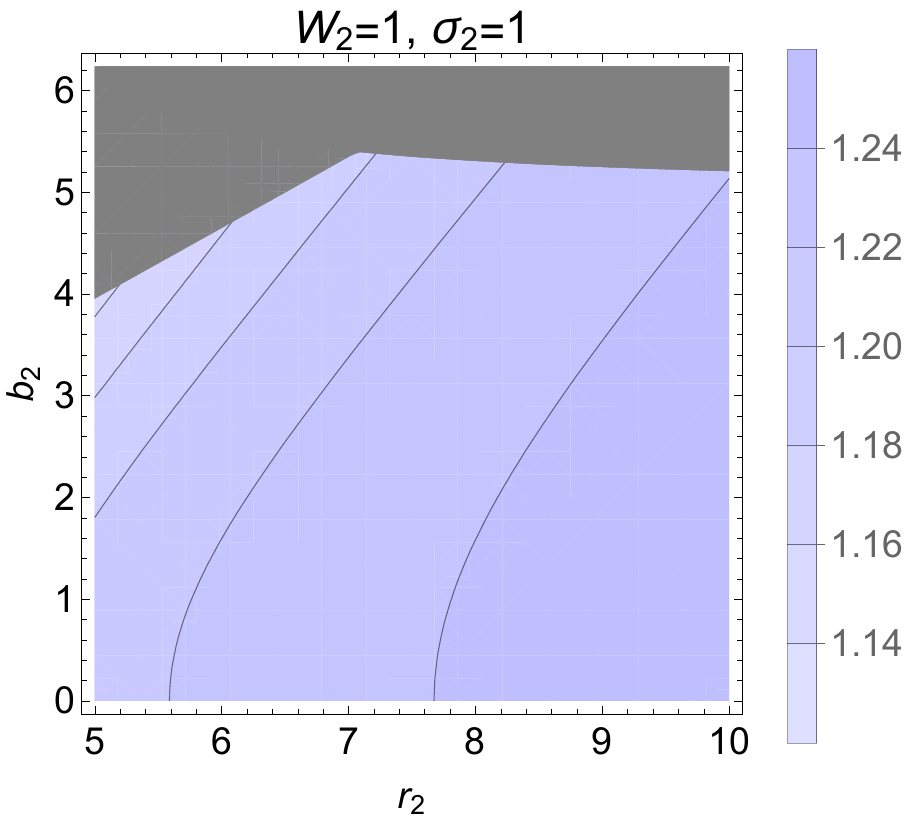}
\caption{\label{fig:expansion-outgoing}
Plot of the redshift factor $\alpha_{12}$ from Eq.~\eqref{eq:redshift-formula} of outgoing ($\sigma_2=+1$) lights from an expanding shell with its velocity, $W_2=0.1,\; 0.5,\; 1$.
The parameters are $r_1=5M$, $W_1=0$, and $\sigma_1=-1$ with the unit $M=1$.
The shaded region is forbidden by the consistency conditions~\eqref{eq:entry-condition}--\eqref{eq:nullvector-condition}.
}
\end{figure}

As a comparison, it is interesting to study a case of expansion.
Here we set $r_1=5M$, $W_1=0$, $r_2>r_1$, $W_2>0$, and $\sigma_1=-1$.
In this case, the light ray should be outgoing at the exit, $\sigma_2=+1$.
The results are shown in Fig.~\ref{fig:expansion-outgoing}.
We can see that the outgoing light rays are definitely blueshifted, $\alpha_{12}>1$, and the larger shell velocity leads to the stronger blueshift.
To summarize the results, Figs.~\ref{fig:collapse-outgoing}--\ref{fig:expansion-outgoing}, there is a clear correspondence between the spacetime dynamics and the redshift of light rays.
The collapse (expansion) of the shell leads to redshift (blueshift) if light rays are outgoing when exiting the shell. The change of the light energy becomes larger for rapid shell dynamics.
Interestingly, an exception occurs when the light rays exit the shell in the ingoing direction. In this case, they can be either redshifted or blueshifted even if the shell is collapsing. We will give this exceptional result a physical interpretation based on covariant analysis in Sec. \ref{sec:generalspacetime}.

\section{Shadow image in gravitational collapse}
\label{sec:shadow-image}
In this section, we demonstrate the shadow formation under several dynamics of the spherical shell.
As inferred from our analysis above, the gravitational collapse of the shell leads to the formation of a black hole shadow due to the redshift of light.
We also see that the shadow disappears if the collapse is halted and an ultracompact object is formed.

We assume that (i) the distant light source is static and distributed homogeneously on a large sphere of radius $r=r_s\gg M$ and light is emitted inwardly from the sphere, (ii) the intensity obeys the Lambert's cosine law, and (iii) the spectrum of the emitted light is monochromatic.
The observer is located at radius $r_o<r_s$ and looking toward the center.
Then, the total energy flux into the observer per unit solid angle and unit area is given as
\begin{align}
\label{eq:energy-flux-formula}
d^2F_{o}=\mathscr{N}\alpha^4\cos\vartheta_o d\Omega_o dS_o,
\end{align}
where $\vartheta_o$ is the incident angle and $\mathscr{N}$ is a factor depending on the radii of the observer and source, the radiance of the source, and so on.
See Appendix~\ref{sec:observation-formalism} for the derivation and details.
The energy flux is basically determined by the fourth power of the redshift factor.
We numerically calculate the energy flux by solving the null geodesic equation backward in time for each observed time $t_o$ and each incident angle $\vartheta_o$ and deriving the redshift factor.

\subsection{Light-speed collapse}
We consider a simple model.
The shell is initially static as $R(\tau)=R_*$ for $t<t_*$ and collapses in the speed of light, implying $ R'(\tau)\to -\infty$, for $t\ge t_*$.
The shell dynamics is implicitly obtained as
\begin{align}
t-t_*=-R+R_*-2M\ln\left[1+\frac{R-R_*}{R_*-2M}\right]
\end{align}
in terms of the Schwarzschild time $t$ from Eq.~\eqref{eq:normalization-sch}, or, 
\begin{align}
T-T_*=-R+R_*
\end{align}
in terms of the Minkowski time $T$ from Eq.~\eqref{eq:normalization-min}.

\begin{figure}[H]
\centering
\includegraphics[width=300pt]{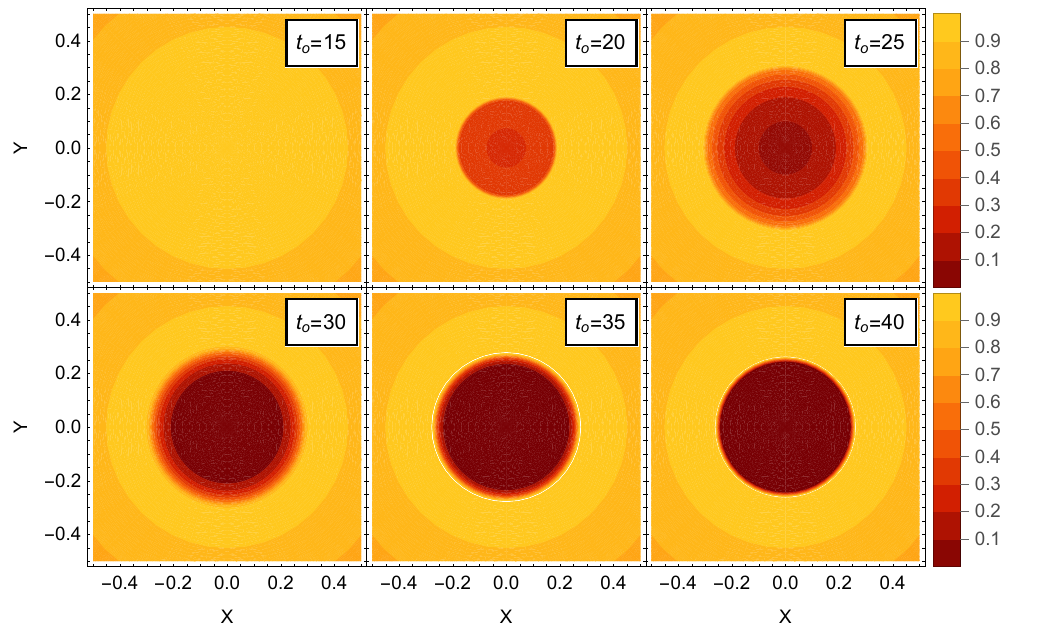}
\caption{\label{fig:nullcollapse-shadow-image}
Snapshots of the shadow image for the light-speed collapse with the normalization $\mathscr{N}=1$.
We set $R_*=5M$, $t_*=0$, $r_\mathrm{o}=20M$, and $r_\mathrm{s}=100M$.
The Cartesian coordinates $X$ and $Y$ are defined on the celestial sphere of the observer so that $X^2+Y^2=\vartheta_o^2$.}
\end{figure}
\begin{figure}[H]
\centering
\includegraphics[width=300pt]{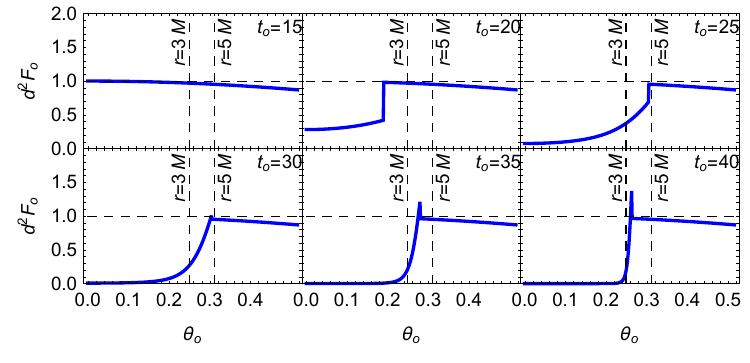}
\caption{\label{fig:nullcollapse-shadow}
The intensity plot for Fig.~\ref{fig:nullcollapse-shadow-image}.
The vertical dashed lines correspond to the incident angles of light rays having turning point at $r=3M$ and $5M$ in Schwarzschild spacetime.
That is, the apparent size of objects with the areal radius $r=3M$ and $5M$.
}
\end{figure}
\begin{figure}[H]
\centering
\includegraphics[width=200pt]{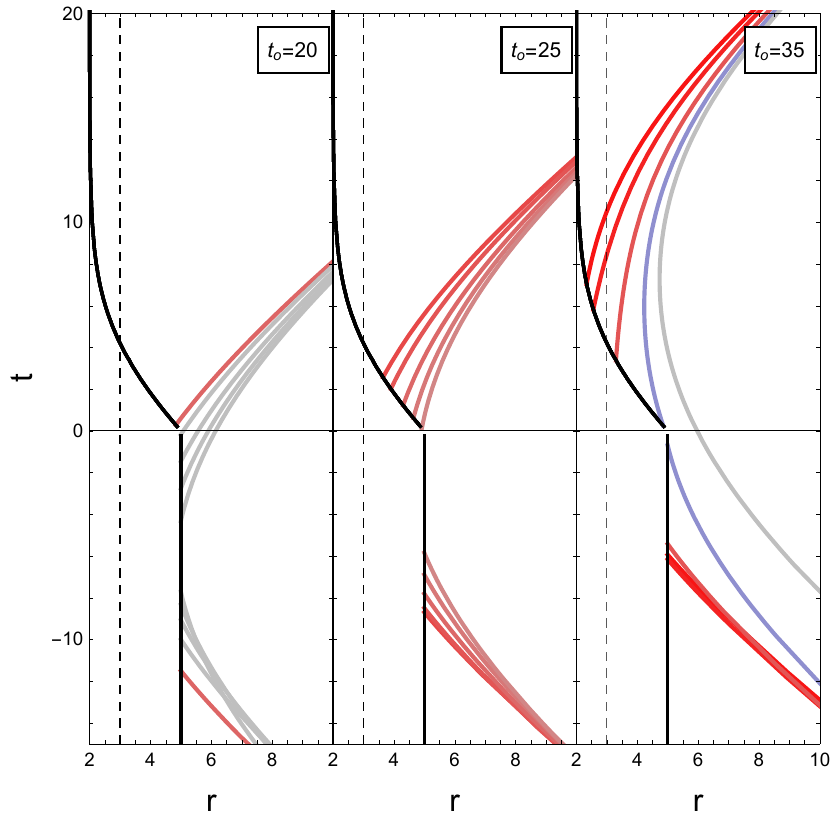}
\caption{\label{fig:nullcollapse-orbits}
Light orbits outside the shell for each observation time for the light-speed collapse.
The incident angles of the orbits are, $\vartheta_o=0.15,\; 0.20,\;0.25, \;0.28$, and $\;0.30$.
The red, blue, and gray orbits correspond to those who are redshifted, blueshifted, and neither.
The black line represents the shell trajectory.
We take the unit such that $M=1$.}
\end{figure}
The results are shown in Figs.~\ref{fig:nullcollapse-shadow-image} and~\ref{fig:nullcollapse-shadow}.
We set $R_*=5M$, $t_*=0$, $r_\mathrm{o}=20M$, and $r_\mathrm{s}=100M$.
For the images in Figure~\ref{fig:nullcollapse-shadow-image}, the distribution of the energy flux $d^2F_o$ is depicted as a color map on the Cartesian coordinates $X,Y$ of the celestial sphere of the observer, which are related to the incident angle as $X^2+Y^2=\vartheta_o^2$.
Fig.~\ref{fig:nullcollapse-shadow} shows the corresponding graph of the flux with respect to $\vartheta_o$.
We adopt the unit $\mathcal{N}=1$ so that $d^2F_o=1$ at $X=Y=0$ before the collapse starts.
We can see that the optical image is getting darker after the collapse starts and finally a shadow is clearly formed.
The evolution can be understood from the light orbits shown in Fig.~\ref{fig:nullcollapse-orbits}.
At the early stage, the central region of the observer's sight becomes darker.
This region corresponds to the incident angle of light rays with small impact parameters and these are the first ones that exit the shell after the collapse starts.
At the next stage, light rays with larger impact parameter also arrive at the observer and the dark region of the image expands.
The dark region expands up to the size corresponding to the initial apparent size of the shell.
At the late stage, light rays with subcritical impact parameter $|b|<b_c$ arrive at the observer with high redshift and light rays with supercritical impact parameter $|b|>b_c$ are no longer redshifted because they are those who have never crossed the shell.
An exception in this case is light rays with super-near-critical impact parameter.
In the images at $t_o=35$ and $40$ of Fig.~\ref{fig:nullcollapse-shadow}, they are blueshifted rather than redshifted at an incident angle slightly outside the photon sphere.
One of them is the light ray with $t_o=35$ and $\vartheta_o=0.28$ shown in Fig.~\ref{fig:nullcollapse-orbits}.
The orbit is ingoing ($\dot r<0$) when exiting the shell and indeed, as suggested from Fig.~\ref{fig:collapse-ingoing}, such an ingoing ray can be blueshifted if the shell is rapidly collapsing.
Since light rays with near-critical impact parameter wind around the photon sphere and stay there for an arbitrarily long time, the blueshifted light rays can be observed even at the later observation time.

\subsection{Dust shell collapse}
Next, we consider more moderate and physically reasonable dynamics.
The shell is initially static as $R(\tau)=R_*$ for $t<t_*$ and starts to collapse as a dust shell for $t\ge t_*$ with zero initial velocity $R'=0$ at $t=t_*$.
The dynamics is given by~\cite{textbook:poisson}
\begin{align}
M=m\sqrt{1+R'^2}-\frac{m^2}{2R},
\end{align}
where the shell's mass $m$ satisfies
\begin{align}
m=R_*\left(1-\sqrt{1-2M/R_*}\right)
\end{align}
so that $R'=0$ at $t=t_*$.

\begin{figure}[H]
\centering
\includegraphics[width=300pt]{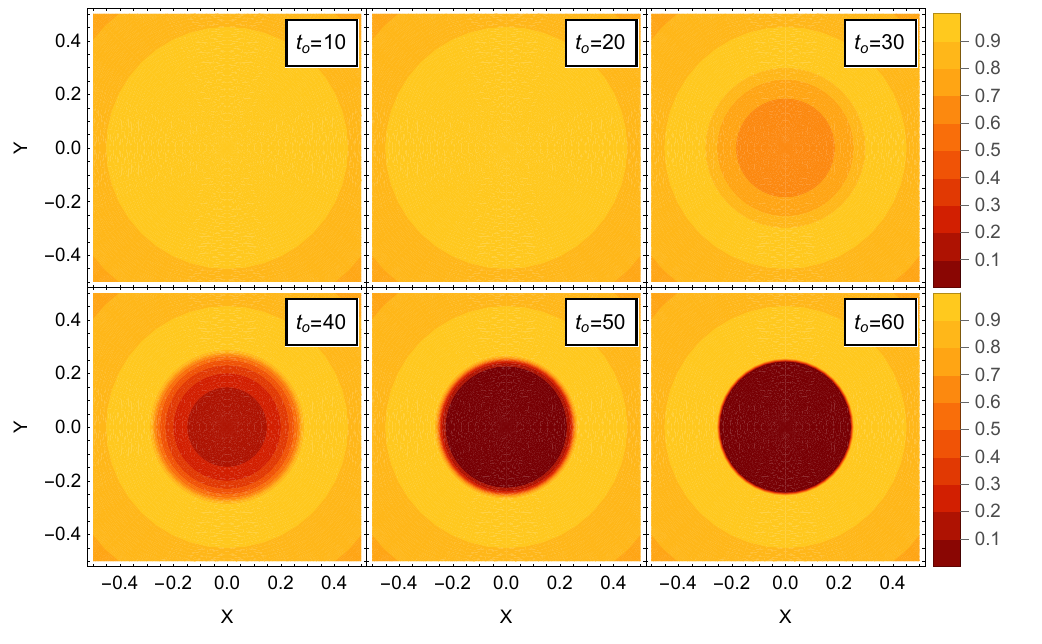}
\caption{\label{fig:dustcollapse-shadow-image}
Snapshots of the shadow image for the dust shell collapse with the normalization $\mathscr{N}=1$.
We set $R_*=5M$, $t_*=0$, $r_\mathrm{o}=20M$, and $r_\mathrm{s}=100M$.
}
\end{figure}
\begin{figure}[H]
\centering
\includegraphics[width=300pt]{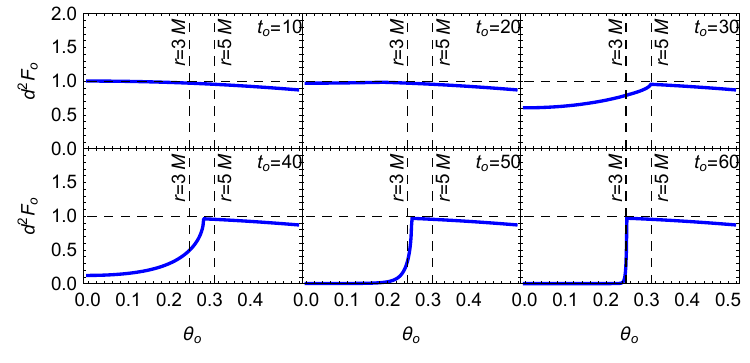}
\caption{\label{fig:dustcollapse-shadow}
The intensity plot for Fig.~\ref{fig:dustcollapse-shadow-image}.
}
\end{figure}
\begin{figure}[H]
\centering
\includegraphics[width=200pt]{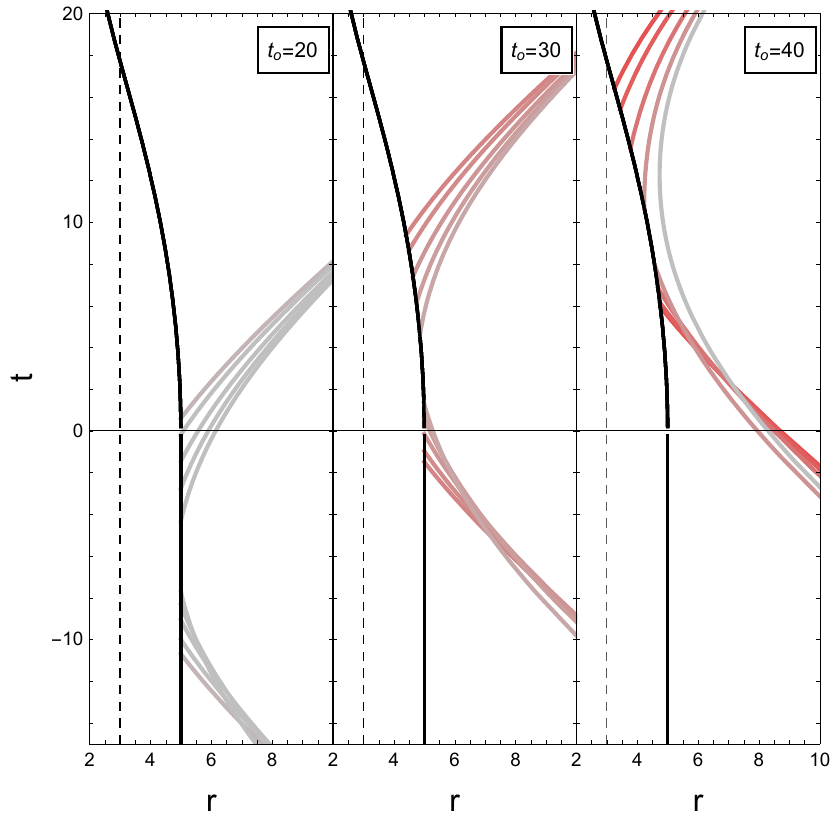}
\caption{\label{fig:dustcollapse-orbits}
Light orbits outside the shell for each observation time for the dust shell collapse.
The incident angles of the orbits are $\vartheta_o=0.15,\; 0.20,\;0.25, \;0.28$, and $\;0.30$.
The red, blue, and gray orbits correspond to those who are redshifted, blueshifted, and neither.
The black line represents the shell trajectory.
}
\end{figure}
The result is shown in Figs.~\ref{fig:dustcollapse-shadow-image} and~\ref{fig:dustcollapse-shadow}.
We set the same parameter values, $R_*=5M$, $t_*=0$, $r_\mathrm{o}=20M$, and $r_\mathrm{s}=100M$ as the previous case.
In this model, the observed image moderately gets darker and finally a shadow is clearly formed with the same apparent size as the photon sphere.
The key difference from the light-speed collapse case is that there is no blueshifted light.
The result is consistent with the analysis shown in Figure.~\ref{fig:collapse-ingoing}, which implies that blueshift occurs only if the shell is rapidly collapsing.
Fig.~\ref{fig:dustcollapse-orbits} shows that indeed the timescale of the shell dynamics is somewhat long compared to the time that light rays are propagating inside the shell.

\subsection{UCO formation}
\begin{figure}[H]
\centering
\includegraphics[width=260pt]{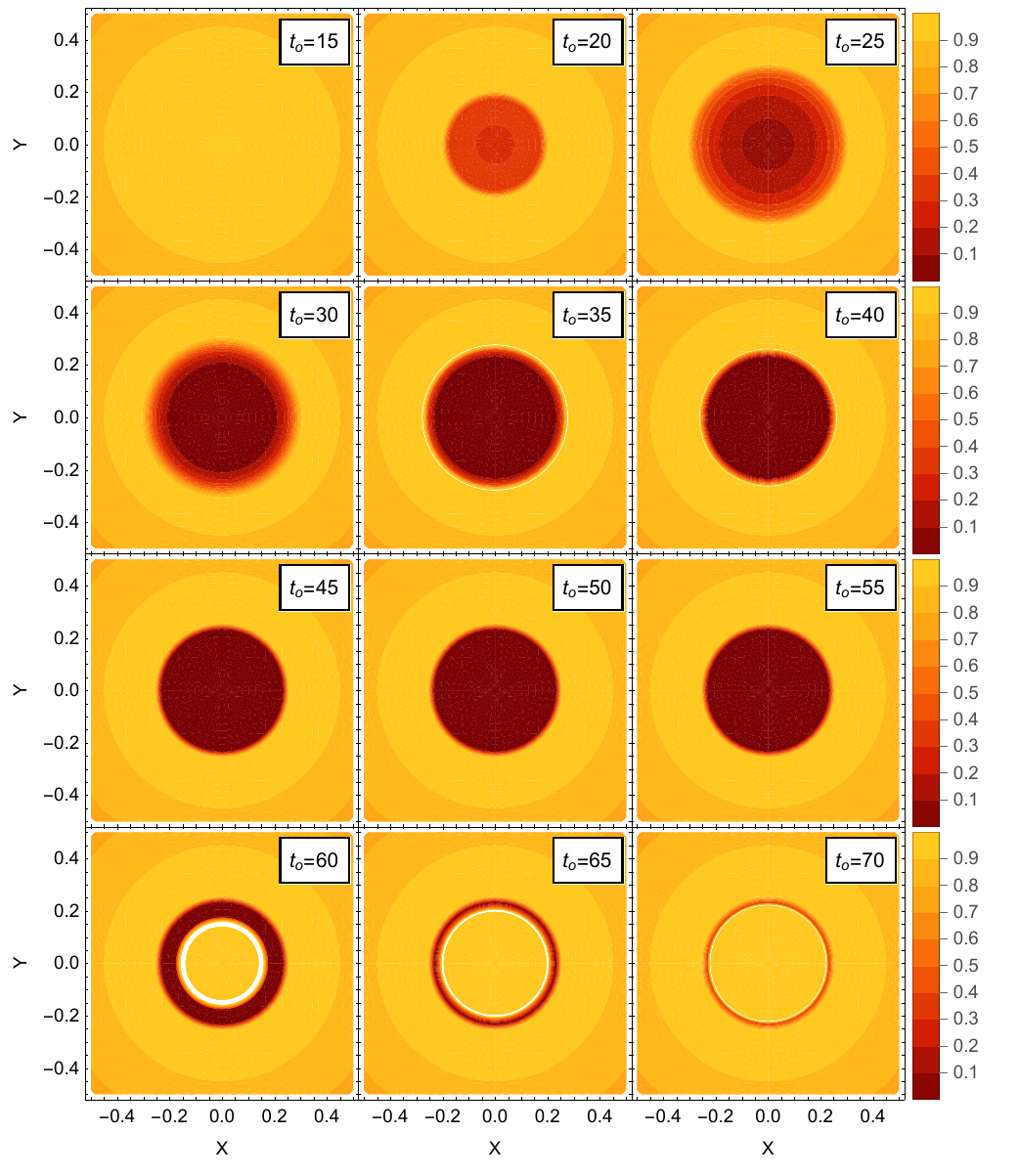}
\caption{\label{fig:nulluco-shadow-image}
Snapshots of the shadow image for the later stage of the UCO formation with the normalization $\mathscr{N}=1$.
We set $R_*=5M$, $t_*=0$, $r_\mathrm{o}=20M$, and $r_\mathrm{s}=100M$.
}
\end{figure}

\begin{figure}[H]
\centering
\includegraphics[width=300pt]{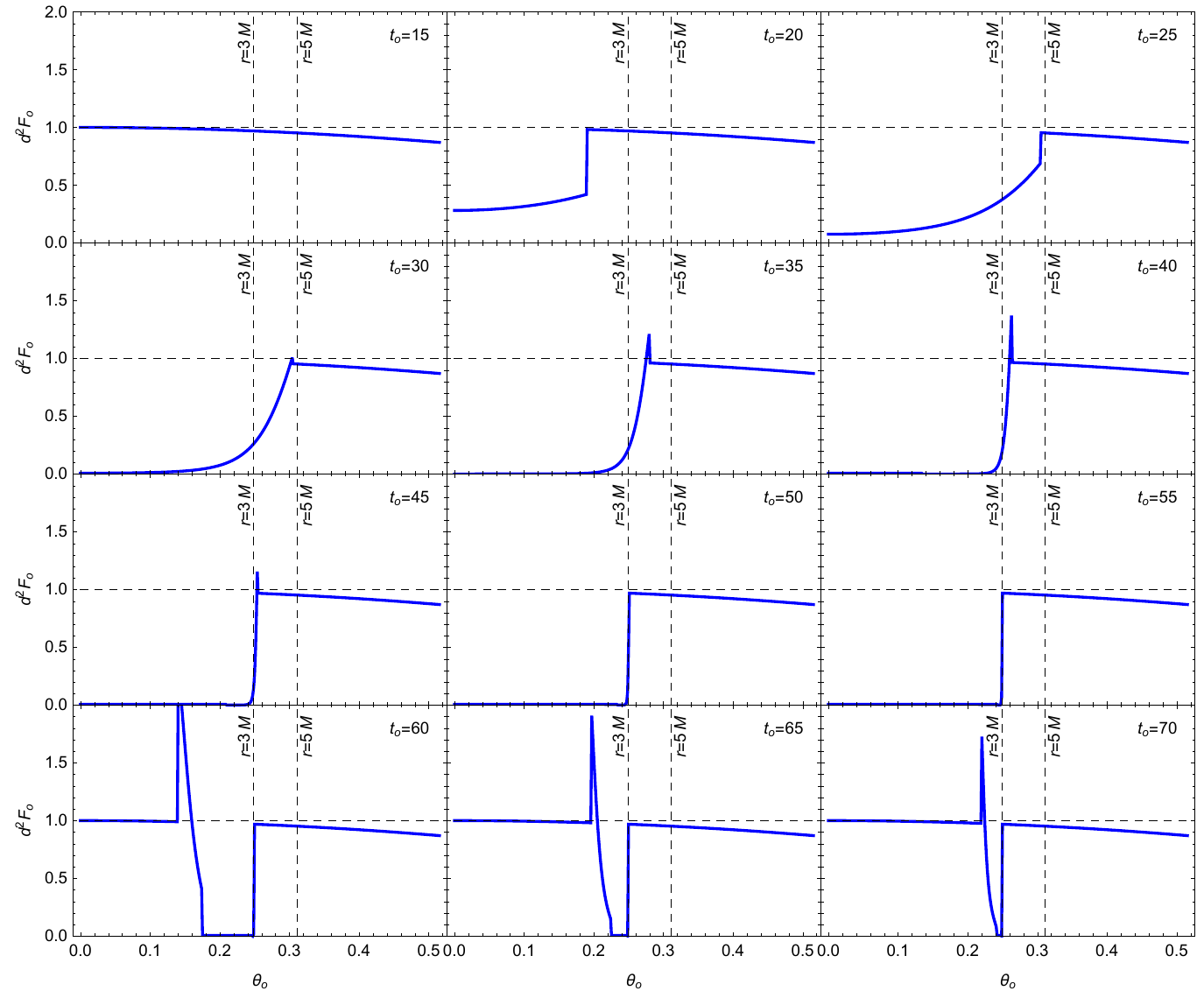}
\caption{\label{fig:nulluco-shadow}
The intensity plot for Fig.~\ref{fig:nulluco-shadow-image}.
}
\end{figure}

\begin{figure}[H]
\centering
\includegraphics[width=200pt]{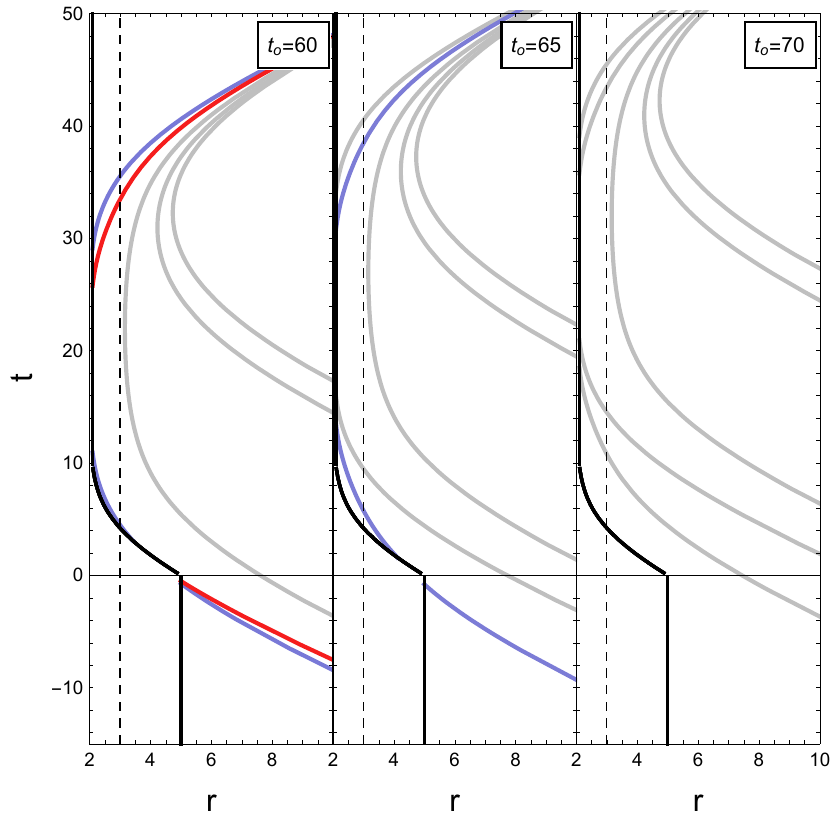}
\caption{\label{fig:nulluco-orbits}
Light orbits outside the shell for each observation time for the UCO formation.
The incident angles of the orbits are, $\vartheta_o=0.15,\; 0.20,\;0.25, \;0.28$, and $\;0.30$.
The red, blue, and gray orbits correspond to those which are redshifted, blueshifted, and neither.
The black line represents the shell trajectory.
}
\end{figure}

Here we suppose a case where the shell collapse is halted at some radius and a static ultracompact object (UCO) is formed.
We assume the same dynamics as the case of light-speed collapse until the shell shrinks to the radius $r=2.1M$.
At the radius, the shell becomes static forever.
The result is the same as Figs.~\ref{fig:nullcollapse-shadow} and~\ref{fig:nullcollapse-orbits} until the shell stops at $t_o\sim 55$.
\footnote{Note that we took fixed resolution $\Delta\vartheta_o\simeq 0.00129$ for the calculation of Fig.~\ref{fig:nulluco-shadow}.
The outer steep peak appearing at $t_o\sim 35$ gets thinner as time evolves and becomes invisible at $t_o\sim 50$. However, we can find the peak surviving at later time with finer resolution.}
After that, the shadow image disappears as shown in Figs.~\ref{fig:nulluco-shadow-image} and~\ref{fig:nulluco-shadow}, and finally a bright image appears.
During the process, at first the center of the observer's sight becomes very bright and this region expands as a bright ring, as clearly seen in Fig.~\ref{fig:nulluco-shadow}.
The bright ring expands toward the apparent size of the photon sphere and finally becomes infinitely thin.
As depicted in Fig.~\ref{fig:nulluco-orbits}, the very bright region corresponds to the blueshifted light rays that are ingoing when exiting the shell during the collapsing phase.
Such light rays are swallowed in a black hole in the case of black hole formation; however, if an UCO is formed, they enter and exit the shell again and eventually escape to infinity.


\section{General spacetime and a covariant formula}
\label{sec:generalspacetime}
In this section, we consider a general, dynamical, asymptotically flat, spherically symmetric spacetime.
Although there is no preferred timelike vector in this spacetime, we take the Kodama vector as the reference of time.
With this choice, we derive a novel covariant formula that represents the redshift of light caused by the spacetime dynamics.

\subsection{Kodama vector and energy of light}
Let $(\mathcal{M},g)$ be a spherically symmetric asymptotically flat spacetime with the metric,
\begin{align}
g=G_{AB}(x^C)dx^Adx^B+r^2(x^C)d\Omega^2,
\end{align}
and $(\mathcal{N},G)$ be its two-dimensional submanifold orthogonal to the sphere of the symmetry.
The Kodama vector~\cite{Kodama:1979vn} is defined as 
\begin{align}
K:=\mathrm{curl} r=-\left(\epsilon^{AB}\partial_B r\right)\partial_A,
\end{align}
where the curl and the totally antisymmetric tensor $\epsilon=2\sqrt{-\mathrm{det}G}dx^0\land dx^1$ are with respect to $(\mathcal{N},G)$ and $x^0$ and $x^1$ are time and radial coordinates, respectively.
This definition ensures that $K$ is future directed.

The Kodama vector was introduced to define a conserved quasilocal mass~\cite{Kodama:1979vn,Hayward:1994bu,Hayward:1997jp}.
It satisfies the two divergence-free equations, $\nabla_a K^a=0$ and $\nabla_a J^a=0$, where $J^a:=-T^a{}_bK^b$ is called the Kodama current and $T_{ab}$ is the energy-momentum tensor defined through the Einstein equation.
The former equation leads to the conservation of the volume $V=-\int_\Sigma d\Sigma g(u,K)$ inside a sphere of radius $r$ on an arbitrary spacelike hypersurface with regular center, $\Sigma$, where $u$ is the unit future-directed normal vector to $\Sigma$.
The latter equation leads to the conservation of the Kodama mass, or equivalently the Misner-Sharp mass, $M_\mathrm{MS}:=r(1-g^{-1}(\nabla r,\nabla r))/2=\int_\Sigma d\Sigma g(u,J)$.
Due to this property, the Kodama vector is sometimes said to be an extension of a timelike Killing vector.
Actually in a static case, one can show that the Kodama vector is parallel to the Killing vector and, in particular, they coincide with each other in vacuum, and thus, in the asymptotically flat region.
\footnote{
More generally, the Kodama vector and a timelike Killing vector agree in electro-$\Lambda$ vacuum, i.e., in the Reissner-Nordstr\"{o}m-(anti-)de Sitter spacetime~\cite{Hayward:1997jp}.
}

Here, we adopt the Kodama vector as a reference timelike vector field to define the energy of a null geodesic:
\begin{align}
\label{eq:K-energy}
E:=-g(k,K),
\end{align}
where $k$ is the geodesic tangent.
It is manifest that the energy coincides with the conserved energy associated with the static Killing vector in the asymptotic region.

\subsection{Covariant redshift formula}
We can consistently reinterpret the definition~\eqref{eq:def-redshift} as the redshift factor defined by the energy associated with the Kodama vector.
Let us express it as
\begin{align}
\alpha
:=\frac{E|_{\mathscr{I}^+}}{E|_{\mathscr{I}^-}}
=\frac{\int_\gamma \nabla_k E d\lambda+E|_{\mathscr{I}^-}}{E|_{\mathscr{I}^-}},
\end{align}
in terms of the energy~\eqref{eq:K-energy}, where $\gamma$ is the null geodesic and $\lambda$ is its affine parameter.
Using the geodesic equation, the derivative of the energy is given as
\begin{align}
\label{eq:nabla-E}
\nabla_kE
=-\nabla_k g(k,K)
=-g(\nabla_k k,K)-g(k,\nabla_k K)
=-g(k,\nabla_k K)
=-\nabla_{(a} K_{b)}k^ak^b,
\end{align}
where the brackets in the indices are for the symmetrization.
For the tensor $\nabla_{(a} K_{b)}$, we have the following proposition.
\begin{proposition}
\label{prop:symd-K}
Let $(\mathcal{M},g)$ be a spherically symmetric spacetime and $(\mathcal{N},G)$ be the two-dimensional submanifold orthogonal to the sphere of the symmetry.
Let $T_{ab}$ be the energy-momentum tensor of $(\mathcal{M},g)$, which satisfies the Einstein equation.
Then, the symmetric derivative of the Kodama vector $K$ satisfies
\begin{align}
\label{eq:symd-K}
\nabla_{(a}K_{b)}=4\pi r \widetilde{\mathcal{T}}_{ab},
\end{align}
where $\widetilde{\mathcal{T}}_{ab}=\epsilon_a{}^c\mathcal{T}_{cb}$ is the dual of $\mathcal{T}_{ab}$, $\mathcal{T}_{ab}={T_\mathcal{N}}_{ab}-\frac{1}{2}\mathrm{tr}_G({T_\mathcal{N}})G_{ab}$ is the trace-free part of ${T_\mathcal{N}}_{ab}$ with respect to $(\mathcal{N},G)$, ${T_\mathcal{N}}_{ab}$ is the restriction of the energy-momentum tensor $T_{ab}$ onto $(\mathcal{N},G)$, and $\epsilon_{ab}$ is the totally antisymmetric tensor on $(\mathcal{N},G)$.
\end{proposition}
One can prove the proposition by applying the Einstein equation to both sides of the equation.
Since the calculation is straightforward, we omit the proof here.
We can see that $\widetilde{\mathcal{T}}=0$ in vacuum and the vanishing $\nabla_{(a} K_{b)}$ implies that $K$ is a static Killing vector, whose existence in a vacuum spherically symmetric spacetime is guaranteed by Birkhoff's theorem.
Note that $\tcal{T}$ is a trace-free symmetric tensor on $(\mathcal{N}, G)$.
One can prove that from the fact that $\mathcal{T}$ is a two-dimensional two-rank tensor which is trace-free and symmetric.

Using Proposition~\ref{prop:symd-K} in the right-hand side of Eq.~\eqref{eq:nabla-E}, we obtain one of our main results:
\begin{align}
\label{eq:redshift-current}
\nabla_k E=-4\pi r \tcal{T}(k,k)
\end{align}
Adopting the Kodama vector as the reference of time, the energy evolution of a light ray is expressed in terms of the energy-momentum tensor. 

Note that the above argument can be also applied to a timelike geodesic.
For a timelike geodesic tangent $\mathfrak{u}$, the energy $\mathscr{E}=-g(\mathfrak{u},K)$ satisfies 
\begin{align}
\label{eq:redshift-current-timelike}
\nabla_\mathfrak{u} \mathscr{E}=-4\pi r\tcal{T}(\mathfrak{u},\mathfrak{u}).
\end{align}

\subsection{Interpretation of the redshift formula}
The formula~\eqref{eq:redshift-current} can be intuitively interpreted as follows.
Using the fact that $\tcal{T}$ is trace-free and symmetric, the energy-momentum tensor on the right-hand side reduces to
\begin{align}
\tcal{T}(k,k)
=\tcal{T}(\bar k,\bar k)
=\mathcal{T}(\bar k,\tilde{\bar k})
=T_\mathcal{N}(\bar k,\tilde{\bar k})
=T(\bar k,\tilde{\bar k})
=T(\bar k,\bar{m}),
\end{align}
where $\bar k\in T_p\mathcal{N}$ is the restriction of $k$ onto the tangent space $T_p\mathcal{N}$, $\tilde{\bar k}=\epsilon_B{}^A{\bar k}^B\partial_A=:\bar m$ is the dual of $\bar k$.
We have used the fact that $\tilde{\bar k}$ is orthogonal to $\bar k$ in the second equality and the fact that $\bar k,\; \tilde{\bar k}\in T_p\mathcal{N}$ in the fourth equality.
Since $\bar k\in T_p\mathcal{N}$ is a null or timelike vector and $\bar m\in T_p\mathcal{N}$ is an outward radial vector orthogonal to $\bar k$, the quantity is the energy current into the sphere of $r$ with respect to the frame $\{\bar k, \bar m\}$.
Then $\delta_k M:=4\pi r^2T(\bar k,\bar m)$ can be interpreted as the mass increase inside the sphere of $r$ that the observer along $\bar k$ effectively feels, and our formula is expressed as
\begin{align}
\label{eq:redshift-newtonian}
\nabla_kE=-\frac{\delta_k M}{r}.
\end{align}
Therefore, the reduction of the energy $E$ at the moment can be interpreted as the reduction of the Newtonian potential energy due to the mass increase inside the sphere of $r$.
The result is quite interesting because the Newtonian analogy holds even in the nonlinear and nonadiabatic regime.

The same interpretation is also possible in the massive particle case.
The formula~\eqref{eq:redshift-current-timelike} for a timelike geodesic tangent $u$, can be rewritten as
\begin{align}
\nabla_\mathfrak{u} \mathscr E
=-\frac{4\pi r^2 T(\bar{\mathfrak{u}},\bar{\mathfrak{m}})}{r}
=:-\frac{\delta_\mathfrak{u}M}{r},
\end{align}
where $\bar{\mathfrak{u}}$ is the restriction of $\mathfrak{u}$ onto $T_p\mathcal{N}$ and 
$\bar{\mathfrak{m}}:=\tilde{\bar{\mathfrak{u}}}=\epsilon_B{}^A\bar{\mathfrak{u}}^B\partial_A\in T_p\mathcal{N}$
is an outward spacelike vector orthogonal to $\bar{\mathfrak{u}}$.
Restoring the gravitational constant $G=1$ and the particle mass $m=1$, which has been implicitly set to unity, we get the more familiar expression,
\begin{align}
\label{eq:redshift-newtonian-massive}
\nabla_\mathfrak{u} \mathscr E= -\frac{Gm\delta_\mathfrak{u} M}{r},
\end{align}
where we renormalized the energy, $\mathscr{E}\to m\mathscr{E}$, as the dimensional energy associated with the 4-momentum $p=mu$.
The expression resembles a reduction of Newtonian potential energy of a massive particle due to the increase of the total mass $\delta_\mathfrak{u} M$ inside the sphere.

Let us remark on how our formula is analogous to the Newtonian case.
Consider a massive particle in a circular orbit in a far region and an accreting mass flow that does not change the background spacetime significantly.
From the formula~\eqref{eq:redshift-newtonian-massive}, the energy associated with the Kodama vector changes according to the effective mass flux $\delta_\mathfrak{u}M=4\pi r^2T(\bar{\mathfrak{u}},\bar{ \mathfrak{m}})$.
As we will see in Eq.~\eqref{eq:msmass-nabla} and below, the effective mass flux deviates from the $\mathfrak{u}$ derivative of the conserved mass, i.e., the Misner-Sharp mass, $\nabla_{\mathfrak{u}}M_\mathrm{MS}=4\pi r^2T(K,\bar{ \mathfrak{m}})$, which should be compatible with the mass in the Newtonian analysis.
The difference is the vectors contracted to the energy-momentum tensor, $\bar{\mathfrak{u}}$ and $K$.
They have the same direction but different norms for the circular particle motion since $\bar{\mathfrak{u}}$ is the projection of $u$ onto the time-radius plane.
With respect to the rest frame, the norm of $\bar{\mathfrak{u}}$ is given as the Lorentz factor due to the motion in the angular direction.
The norm of $K$ is $\sqrt{1-\frac{2M_\mathrm{MS}}{r}}$, where $M_\mathrm{MS}$ is the mass inside $r$.
Both are unity in the leading order if the circular motion of the particle is slow compared to the speed of light and the radius is far from the gravitational radius.
This is the case where the formula~\eqref{eq:redshift-newtonian-massive} approaches the usual Newtonian formula.
However, note that they do not exactly coincide due to our unusual choice of the reference of time, the Kodama vector.

\subsection{The red/blueshift and the spacetime dynamics}
As we have confirmed in the shell case, the dynamics of gravitational collapse typically leads to the redshift of light rather than the blueshift.
Especially from Figs.~\ref{fig:collapse-outgoing} and~\ref{fig:collapse-ingoing}, we can see that (i) every light ray is redshifted if it outwardly exits the collapsing shell while some light rays inwardly exiting are blueshifted, (ii) the redshift effect becomes higher in the rapid collapse of the shell, and (iii) the infinite redshift, $\alpha\to 0$, is achieved for a light ray exiting the shell near the horizon radius, $r\to 2M$.
We can understand this behavior using our formula as follows.

\subsubsection{Redshift of outward light and blueshift of inward light}
\begin{figure}[h]
\centering
\includegraphics[width=150pt]{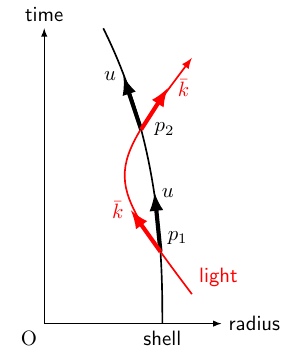}
\caption{\label{fig:light-angle}
The radial velocities of the shell and the light ray during collapse.
}
\end{figure}

Suppose that a light ray crosses a collapsing shell twice as in Fig.~\ref{fig:light-angle}.
Let $p_1$ and $p_2$ be the events where the light enters and exits the shell, respectively.
Since the Kodama vector coincides with the Killing vector in vacuum, the energy of light $E$ defined by the Kodama vector is conserved except for at $p_1$ and $p_2$.
According to our formula~\eqref{eq:redshift-newtonian}, the energy increases when the light enters the shell because $\delta_k M_{p_1}<0$ and decreases when it exits because $\delta_k M_{p_2}>0$.
Intuitively speaking, the value of $\delta_k M$ should be determined by the radial velocity of the light relative to the shell.
Since both the light and shell are ingoing at $p_1$, $|\delta_k M_{p_1}|$ would be small.
On the other hand $|\delta_k M_{p_2}|$ would be large since the light ray is outgoing while the shell is ingoing.
Therefore, we can expect that, in gravitational collapse, the energy increase at $p_2$ is larger than the decrease at $p_1$ leading to the redshift.
We examine this expectation explicitly in the following.

The energy change across the shell is given as follows.
As derived in Appendixes~\ref{sec:shell-em-tensor} or~\ref{sec:sym-d-calculation}, the dual energy-momentum tensor $\tcal{T}$ is given as
\begin{align}
\tcal{T}_{ab}=-\delta(l)\rho u_{(a}n_{b)},
\end{align}
where $\rho:=S(u,u)$ and $S$ is the surface stress tensor given by $T=\delta(l)S$.
From the formula~\eqref{eq:redshift-current}, the energy change $\Delta E$ when the light ray crosses the shell is given by
\begin{align}
\label{eq:DeltaE}
\Delta E
&=\int_{-\delta \lambda}^{\delta \lambda}\nabla_k Ed\lambda \nonumber\\
&=\pm\int_{-\delta l}^{\delta l}-4\pi r \widetilde{\mathcal{T}}(k,k) [g(k,n)]^{-1}dl \nonumber\\
&=\pm\int_{-\delta l}^{\delta l}4\pi r \delta (l)\rho g(u,k)dl \nonumber\\
&=\mp 4\pi R\rho |g(\bar k,u)|,
\end{align}
where the upper (lower) sign is for the event $p_2$ ($p_1$).
$\lambda=0$ is the value of the affine parameter at the shell.
The strength of the red/blueshift depends on $R$, $\rho$, and $|g(k,u)|$, where $R$ is the radius of the shell.
If the shell is collapsing, $R$ decreases while $\rho$ would increase.
The latter contribution would overcome the former since we can expect $\rho\sim R^{-2}$ for a shell.
The factor $|g(\bar k,u)|$ measures the relative radial velocity between the light ray and the shell.
This can be regarded as the Lorentz factor of $\bar k$ in the frame associated with $u$, or vice versa.
It becomes larger when the shell and light ray are moving oppositely and thus, its value is larger at $p_2$.
Consequently, $\Delta E|_{p_1}+\Delta E|_{p_2}<0$ and the light should be finally redshifted, $\alpha<1$.

\begin{figure}[h]
\centering
\includegraphics[width=150pt]{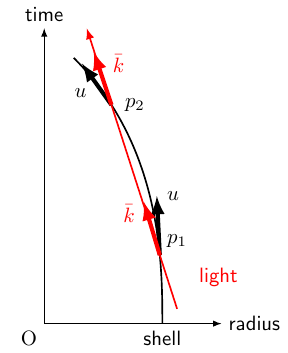}
\caption{\label{fig:light-angle-ingoing}
The ingoing light ray being ingoing at $p_2$.
}
\end{figure}

Let us examine the ingoing light in the collapsing case as depicted in Fig.~\ref{fig:light-angle-ingoing}.
As both $\bar k$ and $u$ are inward, the Lorentz factor $|g(\bar k,u)|$ can become small at $p_2$ resulting in the smaller absolute value of $\Delta E|_{p_2}$.
In total, $\Delta E|_{p_1}+\Delta E|_{p_2}$ can be either positive or negative, i.e., the light can be either blueshifted or redshifted.
In particular, if $\bar k$ is almost aligned to $u$ at $p_2$, $|g(\bar k,u)|$ becomes smaller.
This is why the light with smaller impact parameter $b_2$ tends to be blueshifted in Fig.~\ref{fig:collapse-ingoing}.

\subsubsection{Higher redshift in rapid collapse}
Figure~\ref{fig:collapse-outgoing} shows that, for fixed $b_2$ and $R_2$, the negatively larger $R'$ results in the smaller $\alpha$, where the radial velocity of the shell $R'=dR/d\tau$ is given as the derivative of the radius by the shell's proper time $\tau$.
To understand this redshift behavior for the rapid collapse, Eq.~\eqref{eq:DeltaE} is also useful.
Let us change only the shell velocity $u$ at $p_2$ 
fixing $\bar k$ in Fig.~\ref{fig:light-angle}.
The negatively larger $R'$ corresponds to $u$ tilted away from $\bar k$.
This should imply the relatively larger inward energy flux carried by the shell there.
Actually, $|g(\bar k,u)|$ becomes larger in that case.
Thus, the energy change $\Delta E$ at $p_2$ also becomes negatively larger and the higher redshift, i.e., the smaller $\alpha$, is obtained.

Note that $\rho$ is also dependent on the shell velocity because we are fixing the asymptotic ADM mass, i.e., the total energy of the shell.
From Eq.~\eqref{eq:rho}, we have
\begin{align}
\label{eq:rho-main}
    \rho=\frac{1}{4\pi R}\left(\sqrt{1+R'^2}-\sqrt{f(R)+R'^2}\right),
\end{align}
and this is decreasing in $|R'|$.
On the other hand, for an outgoing light ray when exiting the shell, i.e., $\dot r>0$ at $p_2$ in the Schwarzschild side, we have
\begin{align}
    |g(\bar k,u)|=E_2f(R)^{-1}\left(\sqrt{f(R)+R'^2}-\sqrt{1-b_2^2f(R)R^{-2}}R'\right)
\end{align}
from Eq.~\eqref{eq:shell-4velocity}, where $E_2$ and $b_2=L/E_2$ are the energy and the impact parameter of the light ray, respectively, after exiting the shell.
In the collapsing case, i.e., $R'<0$, this factor is increasing in $|R'|$ with $b_2$ being fixed.
We can numerically check that the product of $\rho$ and $|g(\bar k,u)|$  is increasing in $|R'|$ and thus, the larger Lorentz factor $|g(\bar k,u)|$ dominantly determines the negatively larger $\Delta E$ leading to the higher redshift in the rapid collapse.

\subsubsection{Infinite redshift near the horizon}
Let us make an interpretation for the infinite redshift $\alpha\to 0$ of the light ray near the horizon radius.
For a light ray near the event horizon to escape to infinity, it should be purely radial.
This means that the corresponding null geodesic tangent becomes proportional to the horizon generator in the limiting sense.
Since the horizon in $(M_+,g_+)$ is a Killing horizon, the generator is $\partial_t$ and it coincides with the Kodama vector $K$.
Thus, for the light energy, we have $E=-g(k,K)\propto -g(k,k)=0$ and the infinite redshift $\alpha\to 0$ is inevitably achieved.
This is a brief explanation.
We can also connect the infinite redshift with the formation of the apparent horizon by using our formula as follows.

The apparent horizon formation is signaled by the increase of the Misner-Sharp mass $M_\mathrm{MS}$ inside a radius $r$.
The Misner-Sharp mass is also called the Kodama mass and is given by
\begin{align}
\label{eq:MMS}
    M_\mathrm{MS}=\frac{r}{2}\left(1+g(K,K)\right).
\end{align}
The apparent horizon is formed when the condition $2M_\mathrm{MS}/r=1$ is satisfied~\cite{Hayward:1994bu}.
For the derivative, we can find~\cite{Maeda:2011ii,Burnett:1991nh,Burnett:1993bn}
\begin{align}
\label{eq:msmass-nabla}
    \nabla_aM_\mathrm{MS}=4\pi r^2\epsilon_a{}^c{T_\mathcal{N}}_{cb}K^b.
\end{align}
Then the change of the Misner-Sharp mass along the light trajectory is given by
\begin{align}
    \label{eq:msmass-derivative}
    \nabla_kM_\mathrm{MS}=4\pi r^2 T_\mathcal{N}(\tilde k,K),
\end{align}
where $\tilde k:=\epsilon_B{}^Ak^B \partial_A=\epsilon_B{}^A\bar k^B \partial_A=\tilde{\bar k}$.
This does not coincide with the effective mass flux $\delta_k M=4\pi r^2 T(\bar k,\bar m)=4\pi r^2T_\mathcal{N}(\tilde k,k)$ in our formula~\eqref{eq:redshift-newtonian}.
Obviously, the integration of Eq.~\eqref{eq:msmass-derivative} along the light trajectory gives the difference of the Misner-Sharp mass between the starting point and the endpoint.
In particular, if the starting point and endpoint are in $\mathscr{I}^-$ and $\mathscr{I}^+$, respectively, the integration is trivially zero.
However, the integration of $\delta_kM$ is not trivial and this is why the change in the light energy across the dynamical shell results in the nontrivial redshift despite our Newton-like formula~\eqref{eq:redshift-newtonian}.

Let us examine what is going on for the light ray from almost the horizon radius in terms of $\delta_kM$ and $\nabla_kM_\mathrm{MS}$ in the shell case.
Here we consider a radial outgoing light ray.
The Kodama vector is given as a distribution $K=\Theta(l)K_++\Theta(-l)K_-$, where $K_+=\partial_t$ and $K_-=\partial_T$ are the Kodama vectors on $(M_\pm,g_\pm)$, respectively, and $\Theta(l)$ is {\it the Heaviside distribution} having the value $1$ for $l>0$, $0$ for $l<0$, and indeterminate value at $l=0$~\cite{textbook:poisson}.
From Eq.~\eqref{eq:TN-shell}, the shell energy-momentum tensor projected on $(\mathcal{N},h)$ is given by
\begin{align}
     {T_\mathcal{N}}_{ab}
    =\delta(l)\rho u_{(a} u_{b)}.
\end{align}
For the derivative of the Misner-Sharp mass~\eqref{eq:msmass-derivative}, we have
\begin{align}
\label{eq:nabla-MMS}
    \nabla_kM_\mathrm{MS}
    =4\pi R^2T_\mathcal{N}(\tilde k,K)
    =4\pi R^2\delta(l)\rho g(u,\tilde k)g(u,K).
\end{align}
For the effective mass flux, we have
\begin{align}
    \delta_kM
    =4\pi R^2T_\mathcal{N}(\tilde k,k)
    =4\pi R^2\delta(l)\rho g(u,\tilde k)g(u,k).
\end{align}
They differ only by the factor $g(u,K)$ and $g(u,k)$.
Note that the right-hand side of Eq.~\eqref{eq:nabla-MMS} is not a well-defined quantity as the factor $\delta(l)g(u,K)$ contains the indeterminate product $\delta(l)\Theta(l)$ as a distribution. 
In our case, the delta function and the Heaviside distribution originally come from the matter of the thin shell. Accordingly, the Heaviside distribution should be understood as the limit of a smooth function, and the delta function as the limit of its derivative.
Then we can regard the product as $\delta(l)\Theta(l)\to\frac12\delta(l)$ and
 $\delta(l)g(u,K)$ as $\delta(l)g(u,\frac{1}{2}(K_++K_-))$ in Eq.~\eqref{eq:nabla-MMS}.
\footnote{
From the definition~\eqref{eq:MMS}, we have $M_\mathrm{MS}=\Theta(l)M$ by using the identities $\Theta^2(l)=\Theta(l)$ and $\Theta(l)\Theta(-l)=0$~\cite{textbook:poisson}.
Acting $\nabla_k$ on it,
we get $\nabla_kM_\mathrm{MS}=\delta(l)g(u,\tilde k)M$.
Equation~\eqref{eq:nabla-MMS} agrees with this if the $\delta(l)g(u,K)$ in its right-hand side is regarded as $\delta(l)g(u,\frac{1}{2}(K_++K_-))$.
Equations~\eqref{eq:rho-main} and~\eqref{eq:guK} are also necessary for the check.}
We have
\begin{align}
\label{eq:guK}
    g\left(u,\frac{1}{2}(K_++K_-)\right)
    =\frac{1}{2}\left(g(u,\partial_t)+g(u,\partial_T)\right)
    =-\frac{1}{2}\left(\sqrt{1+R'^2}+\sqrt{f(R)+R'^2}\right),
\end{align}
where we have used Eq.~\eqref{eq:shell-4velocity}.
For the outgoing radial light ray, we have the expression, $k=E(\partial_T+\partial_r^-)$, in the Minkowski side.
Together with Eq.~\eqref{eq:shell-4velocity}, we have
\begin{align}
    g(u,k)=-E(\sqrt{1+R'^2}-R').
\end{align}
In the horizon limit, $R\to 2M$, the above two equations with $R'<0$ give 
$2g(u,\frac{1}{2}(K_++K_-))= g(u,k)/E$
and we obtain
\begin{align}
    \delta_kM/E\to 2\nabla_kM_\mathrm{MS}.
\end{align}
Since the integration of $\nabla_kM_\mathrm{MS}$ across the shell gives the difference of the Misner-Sharp mass $\Delta M_\mathrm{MS}$ between the outside and inside geometries, which is equivalent to the ADM mass in the present case, we can relate the integration of the effective mass flux $\delta_kM$ to the mass as
\begin{align}
\label{eq:DeltakM}
    \Delta_k M:=\int_{-\delta \lambda}^{\delta\lambda} \delta_kM d\lambda
    \to2E\int_{-\delta \lambda}^{\delta\lambda} \nabla_k M_\mathrm{MS}d\lambda
    =2E\Delta M_\mathrm{MS}
    =2EM,
\end{align}
where the integration region $\lambda\in(-\delta\lambda,\delta\lambda)$ is for the affine parameter of the light trajectory across the shell from the inside to the outside.
Equation~\eqref{eq:DeltakM} tells us that, although the horizon is formed if the Misner-Sharp mass $M_\mathrm{MS}$ is confined in the areal radius $R=2M_\mathrm{MS}$, the total effective mass flux injected into the radius, $\Delta_k M$, is the double of it with being normalized by $E$.
Substituting this result into the formula~\eqref{eq:redshift-newtonian}, we have
\begin{align}
\label{eq:DeltaE-horizon}
    \Delta E=\int_{-\delta \lambda}^{\delta\lambda} \nabla_k Ed\lambda
    =-\int_{-\delta \lambda}^{\delta\lambda} \frac{\delta_k M}{R}d\lambda
    =-\frac{\Delta_k M}{R}
    \to-\frac{2\Delta M_\mathrm{MS}}{R}
    =-\frac{2EM}{R}
    =-E.
\end{align}
We can see that the amount of $\Delta_k M$ gives exactly the infinite redshift, $E+\Delta E=0$.

\subsubsection{Kinematical effect on the redshift factor across the shell
}
Decomposing the redshift factor Eq. \eqref{eq:def-redshift} into the sum of the energy change Eq. \eqref{eq:DeltaE}, each energy change includes a gravitational-potential contribution and a kinematical contribution. Usually, these contributions cannot be separated. However, in the shell model, it is possible to partially isolate the kinematical contribution, i.e., the dependence on the radial velocity of the light ray $k$ when the shell's trajectory is fixed. This allows for a more intuitive interpretation of the redshift and blueshift in gravitational collapse, as shown in Figs.~\ref{fig:collapse-ingoing} and~\ref{fig:expansion-outgoing}. 
To this end, we introduce an orthonormal tetrad frame $\{e_{(\mu)}\}$ where the radial velocities of the shell and the light ray are given by $v$ and $w$, respectively, as
\begin{align}
u&=\gamma_{u}\left(e_{(0)}+ve_{(1)}\right), \\
k&=k^{(0)}\left(e_{(0)}+w e_{(1)}+\sqrt{1-w^2}e_{(2)}\right),
\end{align}
with
\begin{align}
    \gamma_{u} =\frac{1}{\sqrt{1-v^2}},
\end{align}
where $\gamma_{u}$ and $\phi_u$ are respectively a Lorentz factor and a hyperbolic angle for the vector $u$, and we assume the light ray moves on the equatorial plane without loss of generality.
Here, it is useful to introduce the normalized vector of $\bar{k}$ as
\begin{align}
    \bar{\bar{k}}&:=\gamma_{\bar{\bar{k}}}\left(e_{(0)}+w e_{(1)}\right)=\frac{\gamma_{\bar{\bar{k}}}}{k^{(0)}}\bar{k}
    \quad \text{with} \quad
    \gamma_{\bar{\bar{k}}}=\frac{1}{\sqrt{1-w^2}}.
\end{align}
By using the normalized vector, we can write the inner product of $k$ and $u$ in terms of the Lorentz factor as
\begin{align}
    g(k,u) = g(\bar k,u) = -\frac{k^{(0)}}{\gamma_{\bar{\bar{k}}}} \gamma_{\bar{\bar{k}}}\gamma_{u}(1-vw)= -k^{(0)}\frac{\gamma_{u\bar{\bar{k}}}}{\gamma_{\bar{\bar{k}}} }.
\end{align}
where $\gamma_{u\bar{\bar{k}} }$ is the relative Lorentz factor of $u$ as seen from $\bar{\bar{k}}$ and given by
\begin{align}
    \gamma_{u\bar{\bar{k}}}:= \gamma_{\bar{\bar{k}}}\gamma_{u}(1-vw) 
    = -g(\bar{\bar{k}}, u).
\end{align}

For the light ray exiting the shell, we take the tetrad frame associated with the Minkowski coordinate basis, $e_{(0)}=\partial_T$, $e_{(1)}=\partial_r^-$, $e_{(2)}=r^{-1}\partial_\theta$, and $e_{(3)}=(r\sin\theta)^{-1}\partial_\phi$.
In this frame, we have $v/\sqrt{1-v^2}=R'$ and $k^{(0)}=E$, where $E$ is the energy evaluated before crossing the shell.
From Eq.~\eqref{eq:rho}, we have
\begin{align}
    \rho=\frac{1}{4\pi R}\frac{1}{\sqrt{1-v^2}}\left(1-\sqrt{f(R)(1-v^2)+v^2}\right),
\end{align}
and then the energy change given by Eq.~\eqref{eq:DeltaE} for the upper sign reduces to
\begin{align}
    \frac{\Delta E}{E}=-\frac{\gamma_{u\bar{\bar{k}}}}{\gamma_{u} \gamma_{\bar{\bar{k}}}} \frac{1-f(R)}{1+\sqrt{f(R)(1-v^2)+v^2}}
    =- \frac{(1-vw)(1-f(R))}{1+\sqrt{f(R)(1-v^2)+v^2}}.
\end{align}
This expression includes kinematical and gravitational-potential contributions that cannot be completely separated. 
However, the dependence on the radial velocity of the light ray $k$ can be extracted as the combination of the Lorentz factors, i.e., $\gamma_{u\bar{\bar{k}}}/(\gamma_{u} \gamma_{\bar{\bar{k}}})=1-vw$, which provides an intuitive interpretation.
When the light ray exits the shell, if both the light ray and the shell asymptotically approach radially null trajectories in the same direction, i.e., $vw \to 1$, 
then the combination of the Lorentz factors becomes zero, and the energy change $\Delta E/E\to 0$.
\footnote{
In such a limit, although $\gamma_{u}$ and $\gamma_{\bar{\bar{k}}}$ diverge, one can check that they cancel out with other factors.
}
In this case, the light ray, which is blueshifted when entering the shell, does not experience much redshift when exiting. As a result, the light ray is blueshifted overall.
In fact, we have already observed such a blueshifted light ray around $b_2\sim 0$ in the right panels of Figs.~\ref{fig:collapse-ingoing} and~\ref{fig:expansion-outgoing}.
On the other hand, if the light ray and the shell asymptotically approach radially null trajectories in the opposite direction, i.e., $v \to -1$ and $w \to 1$, the energy change when exiting reduces to
\begin{align}
    \frac{\Delta E}{E} \to -(1-f(R))\frac{1+w}{2} \to -\frac{2M}{R}.
\end{align}
This means that the light ray is highly redshifted when the shell is highly compactified.
In particular, for $R\to 2M$, the light is infinitely redshifted, $\alpha \to 0$, as discussed previously.

For the light ray entering the shell, we take the tetrad frame associated with the Schwarzschild coordinate basis, $e_{(0)}=f^{-1/2}\partial_t$, $e_{(1)}=f^{1/2}\partial_r^+$, $e_{(2)}=r^{-1}\partial_\theta$, and $e_{(3)}=(r\sin\theta)^{-1}\partial_\phi$.
In this frame, we have $v/\sqrt{1-v^2}=f^{-1/2}R'$ and $k^{(0)}=f^{-1/2}E$.
From Eq.~\eqref{eq:rho}, we have
\begin{align}
    \rho=\frac{1}{4\pi R}\frac{1}{\sqrt{1-v^2}}\left(\sqrt{1-(1-f(R))v^2}-\sqrt{f(R)}\right),
\end{align}
and the energy change given by Eq.~\eqref{eq:DeltaE} for the lower sign reduces to
\begin{align}
    \frac{\Delta E}{E} = \frac{\gamma_{u\bar{\bar{k}}}}{\gamma_{u} \gamma_{\bar{\bar{k}}}}  \frac{f(R)^{-\frac{1}{2}} (1-f(R))}{\sqrt{1-(1-f(R))v^2}+\sqrt{f(R)}}
    =\frac{ f(R)^{-\frac{1}{2}} (1-vw) (1-f(R))}{\sqrt{1-(1-f(R))v^2}+\sqrt{f(R)}}.
\end{align}
Again, the dependence on the radial velocity of the null vector $k$ can be extracted as the same combination of the Lorentz factors, and the energy change $\Delta E/E\to 0$ if both the light ray and the shell asymptotically approach radially null trajectories in the same direction, i.e., $vw \to 1$.
In this case, the light ray has no opportunity to be blueshifted and thus results in an overall redshift.
On the other hand, if the light ray and the shell asymptotically approach radially null trajectories in the opposite direction, i.e., $v \to 1$ and $w \to -1$, we have
\begin{align}
    \frac{\Delta E}{E} \to \frac{1-f(R)}{f(R)} \frac{1-w}{2} \to \frac{1-f(R)}{f(R)}.
\end{align}
This means that the light ray is highly blueshifted when the shell is highly compactified.
In particular, in the horizon limit $R\to 2M$, the light ray is infinitely blueshifted.
This situation corresponds to, for example, an exploding shell emanating from a white hole and a light ray entering into it.

The argument for the single shell provides us an insight into the gravitational collapse of a continuous body.
If the continuous body has negligible radial pressure, it can be regarded as a set of noninteracting cocentric spherical shells.
Collapse of such a continuous body means collapse of each shell and, for each shell, we can expect that most light rays are redshifted after passing through it.
Furthermore, in the late stage of collapse to a black hole, all the light rays escaping to infinity would be redshifted as we have discussed the escapability of blueshifted light rays in Sec.~\ref{sec:collapsing-shell}.
As a consequence, we infer that, even for a continuous body, the dynamics of gravitational collapse quite generally leads to the redshift of observable light rays.
We leave the detailed analysis of this issue as a future work.

\section{Summary and discussion}
\label{sec:summary}
We investigated the redshift of light caused by spacetime dynamics.
Motivated by the black hole shadow formation in gravitational collapse, we focused on light rays emitted and received in the null infinity.
Here, we defined the redshift factor as the ratio of the energy at the future and past null infinity.
If the spacetime is entirely static, the redshift factor is trivially unity.
In this sense, our redshift factor purely reflects the effect of spacetime dynamics.
The detail of our motivation and focus was given in Sec.~\ref{sec:preliminary}.
As a practical example, our configuration, a collapsing object and transmitting light rays, might model a collapsing dark matter and light emissions behind it. 

In Sec.~\ref{sec:collapsing-shell}, we considered a spherical shell model whose dynamics is specified by hand.
We revealed that light rays are typically redshifted if the shell is collapsing, while they are blueshifted if expanding.
If the dynamics is more drastic, i.e., the shell is moving rapidly, the red/blueshift becomes stronger.
The exception is the case where the shell is collapsing and light is radially inward, $\dot r<0$, when exiting the shell.
In this case, the light can be either red or blueshifted.
If the shell finally collapses into a black hole, most of such light rays will be absorbed by the hole.
However, some light rays can escape to infinity if they are reflected by the photon sphere potential after exiting the shell, or if the shell stops collapsing and a horizon is not formed.
Therefore, a distant observer may observe blueshifted light at the early stage of gravitational collapse or after formation of a horizonless compact object.
We have also demonstrated that a shadow is gradually formed during gravitational collapse.
The shadow is clearly shaped at the later stage, however, it disappears if the collapse is halted and a static horizonless compact object is formed.
It is worth noting that the redshift of light in this work is caused by the spacetime dynamics.
Its origin can be distinguished from that of light from a star surface~\cite{Yoshino:2019qsh}, where the light is redshifted by the static gravitational potential and the Doppler shift.

In Sec.~\ref{sec:generalspacetime}, we investigated the redshift of light in a general and dynamical spacetime with spherical symmetry.
To define the energy of light consistently over the whole spacetime, we adopted the Kodama vector as the reference of time.
We found that the evolution of the light energy is then covariantly expressed by the spacetime curvature, or equivalently, the energy-momentum tensor.
This formula allows us to interpret the redshift of light as reduction of the Newton potential due to the effective increase of the total mass inside the radius at the moment.
Finally, we inferred that the spacetime dynamics of gravitational collapse generically leads to the redshift of light when a black hole is likely to form regardless of the detail of the dynamics.

It is interesting to extend the present study to massive particle cases, i.e., the acceleration of massive particles by spacetime dynamics.
As the similar formulas~\eqref{eq:redshift-current-timelike}--\eqref{eq:redshift-newtonian} hold for massive particles, we can expect that the particles may gain their energy in some situations.
Extensive analysis of this problem with various models of collapsing objects may shed light on a new possible mechanism of the astrophysical particle acceleration.
For example, we may apply the analysis to a collapsing ordinary object and cosmic neutrinos passing through it.

The quasilocal interpretation of the change of the energy Eq. \eqref{eq:redshift-newtonian} relies on the property of Kodama vector Eq. \eqref{eq:symd-K}. Regarding the extension of the Kodama vector, the author in \cite{Kinoshita:2024} shows that the Kodama vector or its extension originates from a conformal Killing-Yano two-form. In our case, a two-form $H_{ab}=r(x^A) \epsilon_{ab}$ is indeed a closed conformal Killing-Yano two-form in the four-dimensional spacetime, and the Kodama vector is constructed by $K^{a}=\nabla_{b}H^{ba}/3=-\epsilon^{ab} \nabla_{b}r $. Hence, we expect that a formula similar to Eq. \eqref{eq:redshift-newtonian} could be derived in the spacetime with a conformal Killing-Yano two-form, such as a dynamical and axisymmetric three-dimensional spacetime, and higher-dimensional dynamical warped product spacetimes.

The current work may also have significance in the context of Hawking radiation since the power of such quantum radiation is determined by the redshift factor of corresponding null geodesics \cite{hawking1974, hawking1975}.
Actually, a quantum burst was found in the case of the formation of a horizonless compact object after the gravitational collapse or the vacuum transition in Refs.~\cite{Harada:2018zfg, Barcelo_2019, Kokubu:2019jdx, Okabayashi:2021qfg, Nakao_2022, Nakao_2023}, and the burst implies the rapid time variation of the redshift factor of the corresponding null geodesics. 
Our result for the blueshifted light rays is indeed consistent with the burst phenomena.

In another direction, our redshift formula gives a new insight into the definition of the surface gravity in dynamical spacetime.
A recent proposal is to define it as a quantity that gives the redshift factor of null geodesics at infinity~\cite{Barcelo2011, Barcelo2011_v2, Kinoshita:2011qs}.
However, one may be interested in a definition in terms of a (quasi)local quantity around the gravitating body as its original introduction.
Hayward~\cite{Hayward:1997jp} proposed a notion of {\it dynamical surface gravity} based on a geometrical argument.
He defined it as a coefficient of the {\it antisymmetrized} derivative of the Kodama vector in analogy with the static case.
However, our formula tells us that the redshift of light is relevant to the {\it symmetric} derivative of the Kodama vector.
It is interesting to investigate this apparent discrepancy.

\begin{acknowledgments}
The authors are grateful to T. Harada, C-M. Yoo, S. Kinoshita, N. Oshita, and N. Tanahashi for their fruitful discussions.
The authors also appreciate the referee’s careful reading and thoughtful remarks, which contributed meaningfully to the refinement of the arguments.
This work was supported by JSPS KAKENHI Grants Nos. JP20H05850, JP20H05853, JP21K20367, JP23H01170 (Y. K.), JP22K03626 (M. K.), and JP21J15676, JP23KJ1162 (K.~O.) from the Japan Society for the Promotion of Science.
Y. K. also sincerely thanks C-M. Yoo for providing the opportunity and the support from JSPS KAKENHI Grants Nos. JP20H05850 and JP20H05853, which enabled Y. K. to complete part of this work while working at Nagoya University.
\end{acknowledgments}


\appendix

\section{\MakeUppercase{formalism of the optical observation}}
\label{sec:observation-formalism}
We see how a distant observer observes the brightness around the collapsing object.
The brightness at each point of the observer's celestial sky is the energy flux of the photons into the unit solid angle and unit area.
\par
First we define the number flux of photons per unit solid angle $d\Omega_o$, unit area $dS_o$, and unit frequency $d\omega_o$ into the observer as
\begin{equation}
d^3N_{o,\omega_o}:=I_{o,\omega_o}\cos\vartheta_o d\Omega_o dS_o d\omega_o,
\end{equation}
where $\vartheta_o$ is the angle between the direction of $d\Omega_o$ and the normal to $dS_o$ and $I_{o,\omega_o}$ is the number intensity of photons into the observer's sight within the frequency range from $\omega_o$ to $\omega_o+d\omega_o$.
The number flux of photons to the observer per unit solid angle and unit area is given by
\begin{equation}
d^2N_{o}:=\int d\omega_o \frac{d^3N_{o,\omega_o}}{d\omega_o}=J_o\cos\vartheta_o d\Omega_o dS_o,
\;\;\; J_o:=\int d\omega_o I_{o,\omega_o}.
\end{equation}
Similarly, we define the number flux per unit solid angle $d\Omega_s$, unit area $dS_s$, and unit frequency $d\omega_s$ from the source as
\begin{equation}
d^3N_{s,\omega_s}:=I_{s,\omega_s}\cos\vartheta_s d\Omega_s dS_s d\omega_s,
\end{equation}
where $\vartheta_s$ is the angle between the direction of $d\Omega_s$ and the normal to $dS_s$ and $I_{s,\omega_s}$ is the number intensity of photons emitted from the source within the frequency range from $\omega_s$ to $\omega_s+d\omega_s$.
The number flux of photons from the source per unit solid angle and unit area is given by
\begin{equation}
d^2N_{s}:=\int d\omega_s \frac{d^3N_{s,\omega_s}}{d\omega_s}=J_s\cos\vartheta_s d\Omega_s dS_s,
\;\;\; J_s:=\int d\omega_s I_{s,\omega_s}.
\end{equation}
To relate them, we introduce the source area distance $D_s$ and the observer area distance $D_o$ as
\begin{equation}
\cos \vartheta_odS_o\equiv D_s^2 d\Omega_s,
\;\;\;
\cos \vartheta_sdS_s\equiv D_o^2 d\Omega_o.
\end{equation}
From {\it the reciprocity theorem}, we have
\begin{equation}
D_s^2=D_o^2\alpha_\omega^{-2},
\end{equation}
where $\alpha_\omega:=\omega_o/\omega_s$.
\par
From the conservation of the photon number, we have the equality,
\begin{equation}
d^2N_{s}d\tau_s=d^2N_{o}d\tau_o,
\end{equation}
where $d\tau_s$ and $d\tau_o$ are the proper time intervals of the source and the observer, respectively.
This is the conservation of photons within the tube of the corresponding null geodesic congruence including the width in the time direction.
Then, we have
\begin{eqnarray}
d^2N_{o}&=&J_o\cos\vartheta_o d\Omega_o dS_o\nonumber\\
&=&d^2N_{s}\frac{d\tau_s}{d\tau_o}\nonumber\\
&=&J_s\cos\vartheta_s d\Omega_s dS_s\frac{d\tau_s}{d\tau_o}.
\end{eqnarray}
The observed flux is expressed as
\begin{eqnarray}
J_o&=&J_s\frac{\cos\vartheta_sd\Omega_s dS_s}{\cos\vartheta_o d\Omega_o dS_o}\frac{d\tau_s}{d\tau_o}\nonumber\\
&=&J_s\frac{D_o^2}{D_s^2}\frac{d\tau_s}{d\tau_o}\nonumber\\
&=&J_s\alpha_\omega^2\frac{d\tau_s}{d\tau_o}.
\end{eqnarray}
Using the relation between the time intervals,
\begin{equation}
\omega_sd\tau_s=\omega_od\tau_o,
\end{equation}
or equivalently,
\begin{equation}
\frac{d\tau_s}{d\tau_o}=\frac{\omega_o}{\omega_s}=\alpha_\omega,
\end{equation}
we have
\begin{equation}
J_o=J_s\alpha_\omega^3.
\end{equation}
\par
The energy flux into the observer per unit solid angle, unit area, and unit frequency is given by
\begin{align}
d^3F_{o,\omega_o}
&=\omega_o d^3N_{o,\omega_o}\nonumber\\
&=\omega_oI_{o,\omega_o}\cos\vartheta_o d\Omega_o dS_o d\omega_o\nonumber\\
&=\mathcal{I}_{o,\omega_o}\cos\vartheta_o d\Omega_o dS_o d\omega_o,
\;\;\;\mathcal{I}_{o,\omega_o}:=\omega_o I_{o,\omega_o}
\end{align}
in the unit $\hbar=1$.
The total energy flux per unit solid angle and unit area is given by
\begin{equation}
d^2F_{o}=\int d\omega_o \frac{d^3F_{o}}{d\omega_o}
=\mathcal{J}_{o}\cos\vartheta_o d\Omega_o dS_o,
\;\;\;\mathcal{J}_o:=\int d\omega_o\mathcal{I}_o.
\end{equation}
The energy flux per unit solid angle, unit area, and unit frequency from the source is given by
\begin{equation}
d^3F_{s,\omega_s}=\omega_s d^3N_{s,\omega_s}
=\omega_sI_{s,\omega_s}\cos\vartheta_s d\Omega_s dS_s d\omega_s
=\mathcal{I}_{s,\omega_s}\cos\vartheta_s d\Omega_s dS_s d\omega_s,
\;\;\;\mathcal{I}_{s,\omega_s}:=\omega_sI_{s,\omega_s}.
\end{equation}
The total energy flux per unit solid angle and unit area is given by
\begin{equation}
d^2F_{s}=\int d\omega_s \frac{d^3F_{s}}{d\omega_s}
=\mathcal{J}_{s}\cos\vartheta_s d\Omega_s dS_s,
\;\;\;\mathcal{J}_s:=\int d\omega_s\mathcal{I}_s.
\end{equation}
\par
Here we assume the following light source configuration.
First, the light source is stationary and homogeneously distributed on a large sphere of radius $r_s$.
That is, the spectral photon radiance $I_{s,\omega_s}$ depends on neither the time, $t$, nor the position on the sphere, $(\theta,\phi)$.
The spherical light source surrounds both the central object and the observer and emits light toward its inside.
Second, we impose the Lambert's cosine law, and therefore, $I_{s,\omega_s}$ is also independent of the emission angle $\vartheta_s$.
Third, we assume the monochromatic spectrum with a frequency $\omega_0$.
Consequently, the spectral photon radiance becomes
\begin{equation}
I_{s,\omega_s}=I\delta(\omega_s-\omega_0),
\end{equation}
where $I$ is constant.
\par
We focus our attention on the observable $\mathcal{J}_o$.
The above assumption on the source gives
\begin{equation}
J_s=\int d\omega_s I_{s,\omega_s}=I
\end{equation}
and
\begin{equation}
\mathcal{J}_s=\int d\omega_s \mathcal{I}_s=\int d\omega_s \omega_sI_{s,\omega_s}=\omega_0I=\omega_0J_s.
\end{equation}
Then we have
\begin{equation}
\label{eq:JoJe}
J_o=J_s\alpha_\omega^3=I\alpha_\omega^3.
\end{equation}
From the viewpoint of ray-tracing, a (past-directed) null geodesic in the direction $\vartheta_o$ from an observed point $p_o$ achieves a unique point $p_s$ on the light source.
Along this null geodesic, a frequency $\omega_s$ of a photon at $p_s$ is changed as $\omega_o=\alpha_\omega \omega_s$ with the unique redshift factor $\alpha_\omega$.
Therefore, the monochromatic spectrum of photons at $p_s$ is also observed as monochromatic spectrum at $p_o$.
Then we can express the number intensity as
\begin{equation}
I_{o,\omega_o}=I_o\delta (\omega_o-\alpha_\omega\omega_0)
\end{equation}
with a constant $I_o$.
\footnote{$I_{o,\omega_o}$ can be a function of $\vartheta_o$ depending on the light source configuration in general. Under our assumption, the absence of such dependency is justified by Eq.~(\ref{eq:JoJe})}
Then we have
\begin{align}
J_o=\int d\omega_oI_{o,\omega_o}
=I_o,
\end{align}
and thus,
\begin{equation}
I_o=I\alpha_\omega^3.
\end{equation}
Finally, we have
\begin{eqnarray}
\mathcal{J}_o&=&\int d\omega_o\mathcal{I}_{o,\omega_o}\nonumber\\
&=&\int d\omega_o \omega_oI_{o,\omega_o}\nonumber\\
&=&\int d\omega_o \omega_o\alpha_\omega^3I\delta(\omega_o-\alpha_\omega\omega_0)\nonumber\\
&=&I\alpha_\omega^4\omega_0.
\end{eqnarray}
The energy flux per unit solid angle and unit area is
\begin{equation}
d^2F_{o}=I\alpha_\omega^4\omega_0\cos\vartheta_o d\Omega_o dS_o.
\end{equation}
Letting $k$ be the null geodesic tangent, proper frequency of a light ray, $\omega=-g(k,u_o)$, measured by a static observer $u_o$ is related to the conserved energy, $E=-g(k,\partial_t)$, associated with the timelike Killing vector $\partial_t$ in the asymptotic region as $\omega=\sqrt{f(r)}^{-1}E$.
The redshift factor $\alpha_\omega$ for the frequency is related to the redshift factor $\alpha$ for the asymptotically conserved energy defined in Eq.~\eqref{eq:def-redshift} as $\alpha_\omega=\sqrt{f(r_\mathrm{s})/f(r_\mathrm{o})}\alpha$.
Restoring the unit $\hbar$, the energy flux is finally obtained as
\begin{equation}
d^2F_{o}=\mathscr{N}\alpha^4\cos\vartheta_o d\Omega_o dS_o,\;\; \mathscr{N}=\sqrt{\frac{f(r_\mathrm{s})}{f(r_\mathrm{o})}}I\hbar\omega_0.
\end{equation}
The incident angle $\vartheta_o$ is given as
\begin{align}
    \vartheta_o
    =\left.\tan^{-1}\left(\frac{-r(d\phi)_ak^a}{-\sqrt{g_{rr}}(dr)_ak^a}\right)\right|_{p_o}
    =\tan^{-1}\left(\frac{\sqrt{f(r_o)}}{r_o\sqrt{b^{-2}-f(r_o)r_o^{-2}}}\right)
    \simeq\tan^{-1}\frac{b}{r_o}
\end{align}
in terms of the impact parameter $b$ of the incident light ray, where the last equality holds for a large $r_o$.

\section{\MakeUppercase{Determination of the sign}}
\label{sec:redshift-pm}
The sign $\pm$ in Eq.~\eqref{eq:redshift-formula} is determined as follows.
Let us consider a null geodesic tangent $k$ at $p_2$ in the Schwarzschild side and extend the geodesic backwardly toward the past.
The components are specified by given parameters $E_2$, $b_2$, and $\sigma_2$ as
\begin{align}
    k=f_2^{-1}\partial_t+\sigma_2\sqrt{1-b_2^2f_2r_2^{-2}}\partial_r^+.
\end{align}
The coordinate transformation~\eqref{eq:coordtrans-t} gives the energy in the Minkowski region as
\begin{align}
    \mathcal{E}=k^T=f_2^{-1}\left(A_2-\sigma_2C_2\sqrt{1-b_2^2f_2r_2^{-1}}\right)E_2.
\end{align}
The energy remains constant from $p_2$ to $p_1$.
At $p_1$ in the Minkowski side, we have
\begin{align}
    k=\mathcal{E}\partial_T+\sigma_1^M\sqrt{\mathcal{E}^2-E_2^2b_2^2r_1^{-2}}\partial_r^-,
\end{align}
where we have used Eq.~\eqref{eq:VM} and the fact $L=E_2b_2$.
Here the new parameter $\sigma_1^M=\pm1$ that determines the sign of $k^r_-=\dot r_-$ appears.
By the coordinate transformation $\{\partial_T,\partial_r^-\}\to\{\partial_t,\partial_r^+\}$ at $p_1$, we have
\begin{align}
    E_1=\left(A_1+\sigma_1^MC_1\sqrt{1-\frac{E_2^2b_2^2}{\mathcal{E}^2}r_1^{-2}}\right)\mathcal{E}.
\end{align}
Then, with a little algebra we have
\begin{align}
\label{eq:redshift-sigmaM}
    \alpha_{12}=\frac{E_2}{E_1}
    =\frac{f_2}{f_1}\frac{
A_1\left(A_2-\sigma_2 C_2\sqrt{1-b_2^2f_2 r_2^{-2}}\right)- \sigma_1^MC_1\sqrt{\left(A_2-\sigma_2 C_2\sqrt{1-b_2^2f_2 r_2^{-2}}\right)^2-b_2^2f_2^2r_1^{-2}}
}{
\left(A_2-\sigma_2 C_2\sqrt{1-b_2^2f_2 r_2^{-2}}\right)^2+C_1^2b_2^2f_2^2 f_1^{-1}r_1^{-2}
}.
\end{align}
Thus, the two options of the sign in Eq.~\eqref{eq:redshift-formula} are identified with those of $\sigma_1^M$.

We can relate the parameter $\sigma_1^M$ to $\sigma_1$ in some cases.
Let us go back to the coordinate transformation for $k$ at $p_1$.
For the radial component, we have
\begin{align}
    \dot r_+=C_1\mathcal{E}+A_1\dot r_-,
\end{align}
and thus,
\begin{align}
    \sigma_1=\mathrm{sgn}\left(C_1+A_1\sqrt{1-\frac{E_2^2b_2^2}{\mathcal{E}^2}r_1^{-2}}\sigma_1^M\right).
\end{align}
If the condition,
\begin{align}
\label{eq:sigma-correspondence-condition}
    |C_1|<A_1\sqrt{1-\frac{E_2^2b_2^2}{\mathcal{E}^2}r_1^{-2}},
\end{align}
is satisfied, we have $\sigma_1=\sigma_1^M$.
We can replace $\sigma_1^M$ with $\sigma_1$ in the result~\eqref{eq:redshift-sigmaM} and the redshift factor $\alpha_{12}$ is uniquely determined by specifying the seven parameters $(r_1,r_2,W_1,W_2,\sigma_1,\sigma_2,b_2)$.
So, the minus sign is appropriate in Eq.~\eqref{eq:alpha-e2/e1} in this case.
If the condition~\eqref{eq:sigma-correspondence-condition} is not satisfied, there is not such one-to-one correspondence between $\sigma_1$ and $\sigma_1^M$.
Both choices of $\sigma_1^M=\pm1$ correspond to either $\sigma_1=-1$ or $\sigma_1=+1$.
Specification of the other set of the seven parameters $(r_1,r_2,W_1,W_2,\sigma_1^M,\sigma_2,b_2)$ determines $\alpha_{12}$.
For example, if $r_1=1.01\sqrt{15}M$, $r_2=5M$, $W_1=10$, $W_2=0$, $\sigma_2=+1$, and $b_2=5M$, the condition~\eqref{eq:sigma-correspondence-condition} is violated and $\sigma_1=+1$ for both of $\sigma_1^M=\pm1$.

\section{\MakeUppercase{physical consistency condition for light orbits}}
\label{sec:consistency}
Thanks to the conservation of parameters along a light orbit, we have algebraic equation~\eqref{eq:redshift-formula} of the seven parameters $(r_1,r_2,W_1,W_2,\sigma_1,\sigma_2,b_2)$ for the redshift factor in the shell case.
However, the parameters are restricted by the following three physical consistency conditions.

First, the orbital parameters $r_1,r_2,\sigma_1,\sigma_2,b_2$ should be consistent with the orbit in the interior Minkowski region.
Since the orbit in the Minkowski region is given as the hyperboloid, Eq~\eqref{eq:Min-orbit}, and it has at most one turning point, the radial direction at each point is also restricted.
Let $\sigma^M_i=\mathrm{sgn}(\dot r_-)|_{p_i}$ be the radial direction of the light orbit at each crossing point with respect to the Minkowski coordinates.
If $\sigma^M_1=\sigma^M_2=+1$, the orbit is always outgoing between $p_1$ and $p_2$, and thus $r_2>r_1$.
The case $\sigma^M_1=+1$ and $\sigma^M_2=-1$ is not allowed for the hyperbolic orbit.
In the case $\sigma^M_1=-1$ and $\sigma^M_2=+1$, there is no restriction on the radii.
If $\sigma^M_1=\sigma^M_2=-1$, the orbit is always ingoing between $p_1$ and $p_2$, and thus $r_2<r_1$.
Note that the relation between $\sigma_i$ and $\sigma^M_i$ is read off from Eq.~\eqref{eq:coordtrans-r} as
\begin{align}
\label{eq:sigmaM}
\sigma_i^M=\mathrm{sgn}\left(-f_iC_i+A_i\sqrt{1-b_i^2f_ir_i^{-2}}\sigma_i\right)
\end{align}
and is not trivial for a moving shell with nonzero velocity $W_i$.
Under the above restrictions, the corresponding time of the events $p_1=(t_1,r_1)$ and $p_2=(t_2,r_2)$ are determined by Eq.~\eqref{eq:Min-orbit}.

Second, the velocities of the shell and the light should be taken consistently at each event.
At the entry of a light ray into the shell, the radial velocity of the light ray should be smaller than that of the shell, $dr/dt<dR/dt$, implying that
\begin{align}
\label{eq:entry-condition}
\sigma_1\sqrt{1-b_1^2f_1r_1^{-2}}=
\sigma_1\sqrt{1-\alpha_{12}^2b_2^2f_1r_1^{-2}}<\frac{W_1}{\sqrt{f_1+W_1^2}}.
\end{align}
At the exit, the light velocity should be larger than the shell's velocity, $dr/dt>dR/dt$, implying that 
\begin{align}
\label{eq:exit-condition}
\sigma_2\sqrt{1-b_2^2f_2r_2^{-2}}>\frac{W_2}{\sqrt{f_2+W_2^2}}.
\end{align}

Third, the effective potential $V(E,L;r)$ of a light ray should be negative in Eq.~\eqref{eq:VS}.
This requirement restricts the parameter range by the conditions, $V(E_1,E_1 b_1;r_1)<0$ and $V(E_2,E_2 b_2;r_2)<0$, or equivalently,
\begin{align}
\label{eq:nullvector-condition}
V(1,\alpha_{12} b_2;r_1)=-1+\alpha_{12}^2 b_2^2f(r_1)r_1^{-2}<0,\;\; V(1,b_2;r_2)=-1+b_2^2f(r_2)r_2^{-2}<0.
\end{align}

In summary, the seven parameters $(r_1,r_2,W_1,W_2,\sigma_1,\sigma_2,b_2)$ are restricted by the following conditions:
\begin{itemize}
	\item[(a)] $\sigma_i^M$ given by Eq.~\eqref{eq:sigmaM} must satisfy one of the following:
	\begin{itemize}
		\item[(i)] $\sigma^M_1=\sigma^M_2=+1$ with $r_2>r_1$,
		\item[(ii)] $\sigma^M_1=-1$ and $\sigma^M_2=+1$, or
		\item[(iii)] $\sigma^M_1=\sigma^M_2=-1$ with $r_2<r_1$.
	\end{itemize}
	\item[(b1)] For the light ray to enter the shell at $p_1$, Eq.~\eqref{eq:entry-condition} must hold.
	\item[(b2)] For the light ray to exit the shell at $p_2$, Eq.~\eqref{eq:exit-condition} must hold.
	\item[(c)] The light ray must be in the allowed region $V<0$ of the effective potential in the Schwarzschild region, Eq.~\eqref{eq:nullvector-condition}.
\end{itemize}

\section{\MakeUppercase{Shell's energy-momentum tensor}}
\label{sec:shell-em-tensor}

According to Ref.~\cite{textbook:poisson}, the surface stress energy tensor is given by
\begin{align}
    S_{ab}=-\frac{1}{8\pi}\left([[\mathcal{K}_{ab}]]-[[\mathcal{K}]]h_{ab}\right),
    \nonumber
\end{align}
where the extrinsic curvature of each joined boundary in our case is given as
\begin{align}
    \mathcal{K}_\pm^\tau{}_\tau&=\beta'_\pm/R',\\
    \mathcal{K}_\pm^\theta{}_\theta&=\mathcal{K}_\pm^\phi{}_\phi=\beta_\pm/R,\\
    \beta_+&=\sqrt{f(R)+R'^2},\\
    \beta_-&=\sqrt{1+R'^2}
\end{align}
and the coordinates are taken so that the induced metric becomes
\begin{align}
    h=-d\tau^2+R^2d\Omega^2
\end{align}
for the shell's proper time $\tau$ and the metric of the unit two-sphere, $d\Omega^2=d\theta^2+\sin^2\theta d\phi^2$.
Accordingly, we have
\begin{align}
    S^\tau{}_\tau&=\frac{1}{4\pi R}\left(\beta_+-\beta_-\right),\\
    S^\theta{}_\theta
    &=S^\phi{}_\phi
    =\frac{1}{8\pi}\left(\frac{\beta'_+}{R'}-\frac{\beta'_-}{R'}+\frac{\beta_+}{R}-\frac{\beta_-}{R}\right).
\end{align}

Letting $u$ and $n$ be the 4-velocity of the shell and the outward unit orthogonal vector to the shell, respectively, we have $G_{ab}=-u_au_b+n_an_b$ for the induced metric on the two-surface $\mathcal{N}$.
Then the restriction of the energy-momentum tensor onto $(\mathcal{N},G)$ is given by
\begin{align}
\label{eq:TN-shell}
{T_\mathcal{N}}_{ab}=G_a{}^cG_b{}^dT_{cd}=\delta(l)\rho u_au_b,
\end{align}
where the energy density $\rho$ is given by
\begin{align}
\label{eq:rho}
    \rho:=S(u,u)
    =-S^\tau{}_\tau
    =\frac{1}{4\pi R}\left(\sqrt{1+R'^2}-\sqrt{f(R)+R'^2}\right).
\end{align}
The trace-free part with respect to $(\mathcal{N},G)$ is given by
\begin{align}
    \mathcal{T}_{ab}
    ={T_\mathcal{N}}_{ab}-\frac{1}{2} \mathrm{tr}_G(T_\mathcal{N})G_{ab}
    =\frac{1}{2}\delta(l)\rho \left(u_au_b+n_an_b\right).
\end{align}
Finally, using $\tilde{u}=-n$ and $\tilde{n}=-u$, the dual energy-momentum tensor is
\begin{align}
    \tcal{T}_{ab}=-\frac{1}{2}\delta(l)\rho\left(\tilde u_a u_b+\tilde n_a n_b\right)
    =-\delta(l)\rho u_{(a} n_{b)}.
\end{align}
Substituting Eq.~\eqref{eq:rho}, we have
\begin{align}
    \tcal{T}_{ab}=-\frac{1}{4\pi R}\delta(l)\left(\sqrt{1+R'^2}-\sqrt{f(R)+R'^2}\right)u_{(a} n_{b)}.
\end{align}
We can also derive this result from the symmetric derivative of the Kodama vector by using Proposition~\ref{prop:symd-K} as shown in Appendix~\ref{sec:sym-d-calculation}.

\section{\MakeUppercase{Another derivation of the dual energy-momentum tensor for the shell}}
\label{sec:sym-d-calculation}
Here, we derive the dual energy-momentum tensor $\tcal{T}$ for our shell model from the symmetric derivative of the Kodama vector by using Proposition~\ref{prop:symd-K}.
Letting $K_+=\partial_t$ in the Schwarzschild region and $K_-=\partial_T$ in the Minkowski region, the distribution of the Kodama vector is given as
\begin{align}
    K=\Theta(l)K_++\Theta(-l)K_-.
\end{align}
The radial coordinate $l$ is the affine parameter of the spacelike geodesics perpendicularly penetrating the shell.
The value is taken as $l=0$ on the shell, $l>0$ in the exterior, and $l<0$ in the interior region, and is normalized so that $\nabla_a l=n_a$ for the unit normal vector $n$ to the shell.
The symmetric derivative of the Koadama vector is given by
\begin{align}
\label{eq:sym-d-shell}
\nabla_{(a}K_{b)}
&=[\nabla_{(a}\Theta(l)] {K_+}_{b)}+[\nabla_{(a}\Theta(-l)] {K_-}_{b)}\nonumber\\
&=\delta(l)n_{(a}[[K_{b)}]],
\end{align}
where we have used the Killing equation for the derivative of $K_\pm$ in the first equality.

For the 4-velocity $u$ of the shell and its outward unit normal vector $n$, we have 
\begin{align}
\label{eq:shell-4velocity}
&u=\sqrt{f}^{-1}\sqrt{1+f^{-1}R'^2}\partial_t+R'\partial_r^+
=\sqrt{1+R'^2}\partial_T+R'\partial_r^-,\\
\label{eq:shell-normal}
&n=f^{-1}R'\partial_t+\sqrt{f}\sqrt{1+f^{-1}R'^2}\partial_r^+
=R'\partial_T+\sqrt{1+R'^2}\partial_r^-.
\end{align}
Then we have
\begin{align}
&K_+=\partial_t=-g(\partial_t,u)u+g(\partial_t,n)n
=\sqrt{f}\sqrt{1+f^{-1}R'^2}u-R'n,\\
&K_-=\partial_T=-g(\partial_T,u)u+g(\partial_T,n)n
=\sqrt{1+R'^2}u-R'n,\\
&[[K]]=K_+-K_-=(\sqrt{f}\sqrt{1+f^{-1}R'^2}-\sqrt{1+R'^2})u.
\end{align}
Substituting this equation into Eq.~\eqref{eq:sym-d-shell}, we have
\begin{align}
\nabla_{(a} K_{b)}
=-\delta(l)\left(\sqrt{1+R'^2}-\sqrt{f(r)+R'^2}\right)u_{(a} n_{b)}.
\end{align}
Finally, using Proposition~\ref{prop:symd-K}, we have
\begin{align}
    \tcal{T}_{ab}=\frac{1}{4\pi R}\nabla_{(a} K_{b)}=-\frac{1}{4\pi R}\delta(l)\left(\sqrt{1+R'^2}-\sqrt{f(R)+R'^2}\right)u_{(a} n_{b)}.
\end{align}
The result agrees with that derived from the second junction condition in Appendix~\ref{sec:shell-em-tensor}.



%
\bibliography{collapse_shadow}

\end{document}